\documentclass[useAMS]{mn2e}
\usepackage{epsfig,times,natbib_vw}

\bibpunct[, ]{(}{)}{;}{a}{}{,}
     
\def \aj {AJ}
\def \mnras {MNRAS}
\def \apj {ApJ}
\def \apjs {ApJS}
\def \apjl {ApJL}
\def \aap {A\&A}

\def \araa {ARAA}
\def \pasp {PASP}

\def \physrep {Physics Reports}

\def \logm {$\log({\rm M/M}_\odot)$}
\def \Hdelta {H$\delta$}

\title[Post-starburst galaxies in the VVDS]{Post-starburst galaxies: more than just an interesting curiosity} 
\author[V. Wild et al.]{
\parbox[t]{\textwidth}{\raggedright 
Vivienne Wild$^{1,2}$\thanks{wild@iap.fr},  C. Jakob
Walcher$^2$, Peter H. Johansson$^3$, Laurence Tresse$^4$, St\'{e}phane
Charlot$^2$, Agnieszka Pollo$^{5}$, Olivier Le F\`{e}vre$^4$, Loic
de Ravel$^4$
}
\vspace*{6pt}\\
$^1$Max-Planck Institut f\"{u}r Astrophysik, Karl-Schwarzschild Str. 1,
85741 Garching, Germany \\
$^2$Institut d'Astrophysique de Paris, UMR 7095, 98 bis Bvd Arago, 
75014 Paris, France \\
$^3$Universit\"{a}ts-Sternwarte M\"{u}nchen, Scheinerstr. 1, D-81679
M\"{u}nchen, Germany\\
$^4$Laboratoire d'Astrophysique de Marseille (UMR 6110),
CNRS-Universit\'{e} de Provence, BP 8, 13376 Marseille Cedex 12,
France \\
$^5$The Andrzej Soltan Institute for Nuclear Studies, ul. Hoza 69, 00-681 Warszawa, Poland
}

\begin{document}

\maketitle
\begin{abstract}

From the VIMOS VLT DEEP Survey (VVDS) we select a sample of 16
galaxies with spectra which identify them as having recently undergone
a strong starburst and subsequent fast quenching of star
formation. These post-starburst galaxies lie in the redshift range
$0.5<z<1.0$ with masses $>10^{9.75}M_{\odot}$. They have a number
density of $1\times10^{-4}$ per Mpc$^3$, almost two orders of
magnitude sparser than the full galaxy population with the same mass
limit.
We compare with simulations to show that the galaxies are consistent
with being the descendants of gas rich major mergers. Starburst mass
fractions must be larger than $\sim5-10$\% and decay times shorter
than $\sim10^8$ years for post-starburst spectral signatures to be
observed in the simulations. We find that the presence of black hole
feedback does not greatly affect the evolution of the simulated merger
remnants through the post-starburst phase.
The multiwavelength spectral energy distributions (SEDs) of the
post-starburst galaxies show that 5/16 have completely ceased the
formation of new stars. These 5 galaxies correspond to a mass flux
entering the red-sequence of $\dot{\rho}_{A \rightarrow Q, PSB} =
0.0038^{+0.0004}_{-0.001}$ M$_\odot$/Mpc$^3$/yr, assuming the defining
spectroscopic features are detectable for 0.35\,Gyr. If the galaxies
subsequently remain on the red sequence, this accounts for
$38^{+4}_{-11}$\% of the growth rate of the red sequence. Finally, we
compare our high redshift results with a sample of galaxies with
$0.05<z<0.1$ observed in the Sloan Digital Sky Survey (SDSS) and UKIRT
Infrared Deep Survey (UKIDSS). We find a very strong redshift
evolution: the mass density of strong post-starburst galaxies is 230
times lower at $z\sim0.07$ than at $z\sim0.7$.

\end{abstract}


\section{Introduction}\label{sec:intro}

Since \citet{1926ApJ....64..321H}, the bimodality in the distribution
of galaxy properties has been one of the great curiosities in the
field of astronomy. Bimodality is observed in galaxy colours,
morphology, star formation rates and galaxy stellar masses. Modern
spectroscopic galaxy surveys have allowed us to quantify the
bimodality precisely, especially with the advent of the Sloan Digital
Sky Survey (SDSS)
\citep{2001AJ....122.1861S,2003MNRAS.341...54K,2004ApJ...600..681B}.
Even at high redshift, evidence for bimodality in the galaxy
population is being sought and found
\citep{2005ApJ...625..621B,2006ApJ...647..853W,2007A&A...465..711F}.

Recent observations have revealed that since a redshift of around
unity the total mass of stars living in red sequence galaxies has
increased by a factor of two \citep{2004ApJ...608..752B}. At the same
time, the stellar mass density of the blue sequence has remained
almost constant. The interpretation is that some blue galaxies migrate onto
the red sequence after the quenching of their star formation, whilst
the remainder continue to form new stars 
\citep[e.g.][]{2007ApJ...665..265F,2007A&A...476..137A}. This
quenching of star formation is apparently occurring in galaxies of
increasingly lower masses as the Universe ages
\citep{2006ApJ...651..120B}. So-called ``dry'' (gas poor) mergers,
with little associated star formation, can increase individual galaxy
masses within the red-sequence, forming the massive ellipticals seen
today \citep{2005AJ....130.2647V,2006ApJ...640..241B}. As well as
allowing the building of very massive ellipticals, such a scenario
matches the observed kinematic and photometric properties of
present-day ellipticals
\citep{2006ApJ...636L..81N,2007ApJ...658..710N}. 

The overall decrease in global SFR density since $z\sim1$ appears to
be caused by a gradual decline in the mean SFR of galaxies, rather
than a decrease in the number of starbursts
\citep{1997ApJ...481...49H,2007ApJ...660L..43N,2007A&A...472..403T}. However,
the process, or processes, responsible for this gradual decline remain
to be determined.  What physical mechanisms cause a galaxy to stop
forming stars, to turn into a spheroid, and thus to
enter the red sequence?  There are many competing theories which
address one or both of these problems. Hot gas stripping (sometimes
called strangulation) of small galaxies as they fall into overdense
environments leads to a slow quenching of their star formation
\citep{1980ApJ...237..692L,2000MNRAS.318..703B,2008MNRAS.387...79V}. In
regions of high external pressure, the cold gas may also be stripped
from the galaxies by ram pressure stripping
\citep{1972ApJ...176....1G,1994AJ....107.1003C}. Gas rich major
mergers may perform both tasks, causing a starburst that is strong
enough to rapidly exhaust the supply of fuel, and convert the disk
galaxies into spheroids
\citep{1972ApJ...178..623T,1992ARA&A..30..705B,2003ApJ...597..893N}.
In this latter scenario, subsequent cooling from the inter-galactic
medium may cause a disk to reform and return the galaxy to the blue
sequence. Therefore, the existence and efficient coupling to the ISM
of additional mechanical energy from a central active nucleus (often
called AGN feedback) has been proposed to prevent refueling and allow
the red sequence to build
\citep{2005ApJ...620L..79S,2007ApJ...659..976H,2008MNRAS.387...13K}.
Observational evidence for the existence, but above all the relative
importance of each of these scenarios, remains scarce.

Much progress has been made on understanding galaxy mergers since the
advent of high resolution computer simulations. 
Realistic observational comparisons are now becoming easier through
the application of observational filter functions, point-spread
functions and radiative transfer of starlight through dust, to
simulation outputs \citep[e.g.][]{2008arXiv0805.1246L}. However, many
areas of parameter space remain relatively unexplored with potentially
large implications for our overall understanding of galaxy
formation. For example, the comparison between simulations of galaxy
mergers with and without mechanical AGN feedback is rarely focused upon, with
some groups preferring to focus on starburst feedback, and others on
the possible impact of an active nucleus \citep[for an exception,
see][]{2008MNRAS.387...13K}. 

In this paper we compare observational results directly with a suite
of smoothed-particle hydrodynamic simulations of galaxy mergers from
\citet{2008arXiv0802.0210J}, both with and without AGN feedback, by
converting the star formation histories of the simulated galaxies into
integrated spectra, upon which the same analysis can be performed as
upon the observations. Rest-frame optical spectra contain a wealth of
information about a galaxy's past and present star formation rate,
often referred to as the galaxy ``fossil record''
\citep[e.g.][]{2003MNRAS.343.1145P}. In particular, the 4000\AA\ break
and \Hdelta\ absorption line strengths constrain the specific star
formation rate and amplitude of recent bursts of star formation
\citep{2003MNRAS.341...33K}. At low redshift, the quality of spectra
and models now allow the analysis of large spectral regions
\citep{2005MNRAS.358..363C, 2007MNRAS.tmp..840T, wild_psb,
2008MNRAS.385.1998K}. At high redshift, spectral indices easily
separate blue from red sequence galaxies without the complication of
dust \citep{2008A&A...487...89V}. At these redshifts, where galaxy
spectra are in general noisier, modern statistical techniques present
us with the opportunity to vastly increase the quantity and improve
the quality of derived parameters. Following \citet{wild_psb}, this
paper presents a principal component analysis of the 4000\AA\ break
region of more than one thousand high redshift galaxy spectra culled
from the VIMOS VLT Deep Survey
\citep[VVDS,][]{2005A&A...439..845L}. We use the information encoded
in the spectra to recover the recent star formation history of galaxies
at an epoch when the Universe was very different from today.

From an observational perspective, so-called ``transition'' galaxies
are of significant interest for pinning down which physical processes
play the largest role in shaping the evolution of galaxies.  Classes
of galaxies which could be transitioning between the blue and red
sequences are ``post-starburst'' galaxies
\citep[e.g.][]{2004ApJ...609..683T,
2004ApJ...602..190Q,2006ApJ...642...48L,2007MNRAS.382..960K} and
``green-valley'' galaxies \citep[e.g.][]{2007ApJS..173..342M}. From
the build up of the red sequence in the VVDS,
\citet{2007A&A...476..137A} measured the net mass flux which has taken
place from the blue sequence to the red sequence. This amounts to
$9.8\times10^{-3}$M$_\odot$/yr/Mpc$^3$ for a
\citet{2003PASP..115..763C} initial mass function (IMF), or about
$1.4\times10^4$\,M$_\odot$/yr in the entire VVDS survey volume. As we
shall show in this paper, galaxies which have undergone sudden
quenching of star formation can be identified in VVDS spectra for a
period of $\sim0.35-0.6$\,Gyr after the quenching event. Therefore, if
the entire build up of the red sequence were due to a physical process
associated with fast quenching, such as gas rich major mergers, we
could expect to identify a net transitioning stellar mass of
$6-10\times10^5$M$_\odot$/Mpc$^3$, or about
$8.5-14\times10^{11}$M$_\odot$ in the VVDS survey, after accounting
for targeting rates. This could comprise, for example, a few tens of
galaxies of stellar mass \logm$=10.5$.

The primary questions that we wish to address in this paper are, how
much mass do we see entering the red sequence after an episode of fast
quenching?  Are gas-rich major mergers an important mechanism for the
build up of the red-sequence since $z\sim1$?  Section 2 describes the
VVDS spectroscopic survey and Section 3 presents our method for
identifying post-starburst galaxies. In Section 4 we place our
empirical results within a theoretical framework, through comparisons
with simple synthesised stellar populations and the star formation
histories of galaxy merger simulations. The results are presented in
Section 5 and compared to the low redshift Universe in Section 6. In
Section 7 we discuss the global importance of post-starburst galaxies
for developing a complete picture of the evolution of the galaxy
population.

Throughout the paper we assume a cosmology with $H_0 =
70$\,km\,s$^{-1}$\,Mpc$^{-1}$, $\Omega_{\rm m} = 0.3$ and
$\Omega_\Lambda= 0.7$.

\section{The VVDS dataset}

\begin{figure}
\includegraphics[scale=0.4]{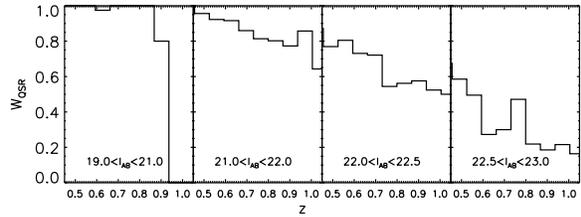}
\caption{The ``quality sampling rate'' between $0.5<z<1.0$ for a SNR limit
  of 6.5 and in four bins of I$_{AB}$ apparent magnitude.}\label{fig:qsr}
\end{figure}

The VIMOS VLT Deep survey (VVDS) is a deep spectroscopic redshift
survey, targeting objects with apparent magnitudes in the range of
$17.5\le I_{AB}\le24$. The survey is unique for high redshift galaxy
surveys in having applied no further colour cuts to minimise
contamination from stars, yielding a particularly simple selection
function. In this work we make use of $\sim9000$ spectra from the
publicly available first epoch public data release of the VVDS-0226-04
(VVDS-02h) field which covers an area of 1750 square arcmins
\citep{2005A&A...439..845L}. The spectra have a resolution (R) of 227
and a dispersion of 7.14\,\AA/pixel and a useful observed frame
wavelength range, for our purposes, of 5500-8500\,\AA.

The first epoch public data release contains 8981 spectroscopically
observed objects in the VVDS-02h field, 8893 of which are within the
presently defined survey mask. Of these, 7194 have secure redshifts
(flags 2, 3, 4 and 9, see below), are not classified as type 1 AGN and
are not secondary objects in the slit. To measure each galaxy's star
formation history we require coverage of the rest-frame wavelength
range 3750--4140\AA, which restricts the redshift range of our galaxy
sample to $0.5<z<1.0$. As discussed in detail in Section \ref{sec:qsr}
below, we impose a signal-to-noise (SNR) limit on the spectra greater
than 6.5 and magnitude limits of $19<I_{AB}<23$, resulting in a final
sample of 1246 galaxies.

In addition to the spectra the dataset includes multi-wavelength
photometry from diverse sources. Data assembled within the context of
the VVDS survey include deep B, V, R, and I photometry from the
CFHT/CFH12K camera\footnote{http://cencosw.oamp.fr/} with limiting
magnitudes of 26.5, 26.2, 25.9, and 25.0, respectively
\citep{2003A&A...410...17M,2004A&A...417..839L}.  Further optical
photometry from the Canada-France-Hawaii Telescope Legacy Survey
(CFHTLS\footnote{http://www.cfht.hawaii.edu/Science/CFHTLS}) extends
into the $u$ and $z$ bands.  The field has deep NIR imaging down to
limits of J$\approx$ 21.50 and K$\approx$ 20.75
\citep{2005A&A...442..423I} and NUV and FUV coverage as part of the
GALEX mission \citep{2005ApJ...619L...1M}. In order to avoid source
confusion due to the large point spread function (PSF) of GALEX, UV
photometry is based upon optical priors (Arnouts et al., in
prep.). Infrared photometry in the IRAC 3.6, 4.5, 5.8 and 8.0 $\mu$m
is available from the Spitzer Wide-area InfraRed Extragalactic survey
\citep[SWIRE,][]{2003PASP..115..897L}. We use the source catalogue as
released by the SWIRE
team\footnote{http://swire.ipac.caltech.edu/swire/swire.html}, with
detection limits at 5$\sigma$ of 5.0, 9.0, 43.0, 40.0 $\mu$Jy,
respectively. In this paper the photometric data is used primarily for
the purpose of deriving stellar masses, as described in \citet[][see
Section \ref{sec:Ms}]{2008arXiv0807.4636W}.

\subsection{Accounting for unobserved galaxies}

In order to calculate absolute quantities from galaxy redshift
surveys, such as galaxy numbers and mass densities, it is necessary to
account for several effects which cause galaxies not to be
included. For our work there are three effects:
\begin{itemize}
\item The {\it target sampling rate} (TSR): only a fraction of all
  galaxies in the given magnitude range and survey area are targeted
  for spectroscopic follow-up.
\item The {\it spectroscopic success rate} (SSR): a fraction of
  targeted and observed galaxies cannot be successfully assigned
  redshifts, due to either sky line contamination or featureless
  spectra.
\item The spectroscopic {\it quality success rate} (QSR): only a
  fraction of the targeted galaxies with successful redshifts have
  high enough quality spectra from which star formation histories
  can be measured.
\end{itemize}
The TSR and SSR are common to many results derived from the VVDS, such
as luminosity functions, and are explained in detail in
e.g. \citet{2005A&A...439..863I}; only the QSR is new to this
work.

\subsubsection{The Target Sampling Rate (TSR)}

As described in \citet{2005A&A...439..863I}, in order to maximise the
number of targeted objects per VIMOS pointing, the TSR is a function
of the size of the galaxy. We obtain the TSR from Figure 1 of
\citet{2005A&A...439..863I} and weight each galaxy according to its
size $w_i^{\rm TSR} = 1/{\rm TSR(r_i)}$, where $r_i$ is the x-radius
parameter of each galaxy.

\subsubsection{The Spectroscopic Success Rate (SSR)}

The second effect, the SSR, is potentially much more complicated,
because the success of obtaining a redshift for a galaxy depends not
only on the apparent magnitude of the galaxy, but also on its redshift
and spectral type.  The rest-frame spectroscopic features which allow
accurate redshift determinations depend on the spectral type of the
galaxy and move through the observed-frame wavelength range depending
on the redshift of the galaxy. In the VVDS each object is assigned a
redshift quality flag, which approximates the security of the given
redshift. Galaxies with flags 2, 3, 4 and 9 are deemed
secure\footnote{Flag 9: redshift determined from single emission line;
Flag 4: 100\% confidence; Flag 3: 95\% confidence; Flag 2: 75\%
confidence \citep{2005A&A...439..863I}}. In order to account for the
changing SSR as a function of redshift, and following the method
presented in Section 3.2 of \citet{2005A&A...439..863I}, we estimate
the SSR by making use of the photometric redshifts calculated by
\citet{2006A&A...457..841I}, which are publicly available through the Terapix
Consortium\footnote{ftp://ftpix.iap.fr/pub/CFHTLS-zphot-v1/}. We
calculate the SSR as a function of redshift and magnitude, as the
number of objects with secure redshifts (flags 2,3,4,9) divided by the
total number of objects (with flags 0,1,2,3,4 and 9).  Photometric
redshifts are used to place the galaxies without secure spectroscopic
redshifts in the correct redshift bin.

We calculate the SSR in 4 bins of apparent magnitude
19--21, 21--22, 22--22.5 and 22.5--23, with bins in $\Delta
z_{\rm phot}$ of width 0.05. Galaxies brighter than 19 and
fainter than 23 magnitudes are not used in this work, the former
because of our lower redshift limit of $z=0.5$ and the latter because
of the signal-to-noise limitation we place on the spectra as described
in the following subsection.

We weight each galaxy in our sample by $w_i^{\rm SSR}= 1/{\rm
SSR(z_{spec},} I_{AB})$, where $z_{spec}$ is the spectroscopic redshift
of the galaxy, and $I_{AB}$ is its $I$-band apparent magnitude.

It is important that galaxies with different SEDs are treated equally
during this weighting procedure. Quiescent galaxies are the hardest to
assign redshifts to, due to their lack of emission lines. To check
that the completeness for quiescent galaxies is accurately estimated
using the SED-independent method described above, we split our sample
into high and low SFR subsamples, where SFR was measured from SED
fitting to multiwavelength photometry (see Section \ref{sec:Ms}). We
repeated the calculation of $w_i^{\rm SSR}$ for each subsample and
found them to be consistent with each other over the redshift range
$0.5<z<1$ used in this paper. This redshift range is expected to be
the least problematic, due to the coverage of the 4000\AA\ break which
is strong in quiescent populations.

\subsubsection{The Quality Success Rate (QSR)}\label{sec:qsr}

The final selection criterium applied to the VVDS galaxies to create the sample
studied in this paper is that they have spectra of high enough
continuum SNR to allow us to extract useful information on their star
formation histories. In practice, this means a median
per-pixel-SNR\footnote{Per-pixel SNR is calculated from the median
flux over error in the wavelength range 5500--8500\AA.} larger than
6.5. The SNR of the observed spectrum depends on both the apparent
magnitude of the galaxy, because all galaxies are observed with equal
exposure times, and the redshift of the galaxy, because the flux of a
galaxy varies with wavelength. 

Thus, in a similar manner as for the SSR, we define the quality
success rate (QSR) to be the total number of galaxies with secure
redshifts and SNR greater than our limit, divided by the total number
of galaxies with secure redshifts (flags 2,3,4 and 9). We calculate
this in the same four bins of apparent magnitude used for the SSR and
in spectroscopic redshift bins of width $\Delta z_{\rm spec}=0.05$. We
weight each galaxy in our sample by $w_i^{\rm QSR}= 1/{\rm
QSR(z_{spec},} I_{AB})$. The $w^{\rm QSR}$ is shown in Figure
\ref{fig:qsr} for the four magnitude bins. In Section
\ref{sec:massdens} we will show that we recover stellar mass densities
consistent with previous work, despite the addition of this new
selection criterion.

Imposing a SNR threshold on the data was found to be necessary for
this work. Using spectra with too low SNR causes considerable scatter
in our spectroscopic indices, which makes difficult the robust
identification of the post-starburst galaxies in which we are
primarily interested. This problem is easily identifiable: the
increased scatter in the indices causes the number of post-starburst
galaxies to increase relative to the total sample, and biases the SNR
distribution of the candidate post-starbursts towards lower values
than typically seen in the population as a whole. Therefore the SNR
limit was chosen empirically, by ensuring that the distribution of SNR
in our post-starburst galaxy sample was similar to that for galaxies
of other types, and that our total completeness-corrected stellar mass
densities for each class of galaxy (i.e. quiescent, star forming,
post-starburst, starburst) remained constant when the SNR limit was
increased by a small amount. We found that a per-pixel-SNR limit of 6.5
was required for the purposes of our study.

\subsection{Measuring Stellar masses}\label{sec:Ms}

A full description of the method used to measure the stellar masses of
our galaxies is given in \citet{2008arXiv0807.4636W}, to which we refer the
reader for details. Briefly, we derive the stellar mass M$^*$ for each
galaxy by fitting its multi-band UV--IR spectral energy distribution
(SED), to model stellar populations created using the
\citep{2003MNRAS.344.1000B} population synthesis models, assuming a
\citet{2003PASP..115..763C} initial mass function (IMF). We create a
library of exponentially declining star formation histories with
superposed stochastic top-hat bursts (see Kauffmann et al. 2003 and
Salim et al. 2005 for descriptions of similar
libraries)\nocite{2003MNRAS.341...33K}\nocite{2005ApJ...619L..39S}.
The method uses Bayesian probability distribution functions as
described in Appendix A of Kauffmann et al. (2003) to derive physical
parameter estimates for each observed SED, including errors on the
parameters.  For a detailed discussion of the accuracy of stellar mass
estimates using SED fitting methods see \citet{2007A&A...474..443P}.

\subsection{Volume correction}\label{sec:volume}

To account for the changing effective survey volume for galaxies of
different brightnesses and SED shapes, we use the V$_{\rm max}$ method
\citep{1968ApJ...151..393S}. This weights galaxies according to the
fraction of the survey volume in which they could have been observed,
given their brightness and the shape of their SED. The latter point is
important: galaxies with different SED shapes are visible in different
survey volumes and this must be correctly accounted for in the V$_{\rm
max}$ calculation \citep{2004MNRAS.351..541I}. Using the same library
of models as used in the stellar mass determination (see above), the
best fit model was scaled to the observed galaxy magnitude and shifted
in redshift to determine the total redshift path in which the galaxy
would have been seen, given the survey magnitude limits.

The SNR limit alters slightly the effective $I_{AB}$ magnitude limits
of the survey, because very few galaxies with apparent magnitudes
above 23 have spectra with a SNR greater than our limit of 6.5. We
therefore redefine the survey magnitude range to be $19<I_{AB}<23$
and remove galaxies from our sample which fall outside of these
limits.

\subsection{The survey volume}
The survey area is calculated from the photometric survey mask,
combined with the boundaries of the spectroscopic survey region. The
total survey area is 0.472 sq. degrees which corresponds to a survey
volume in the redshift range $0.5<z<1.0$ of $1.4\times 10^6$\,Mpc$^3$.



\section{Locating recently dead galaxies}\label{sec:dead}

The galaxies in which we are interested are those in which a
dominant A/F star population is visible, implying that the formation
of O and early B type stars has ceased suddenly. Such galaxies are
identifiable by their strong Balmer absorption lines compared to their
mean stellar age as measured by their 4000\AA\ break strength.  The
traditional method for identifying post-starburst galaxies involves
the selection of objects with both a strong Balmer absorption line
(usually H$\beta$ or H$\delta$) to identify a recent starburst, and no
detectable nebular emission (usually [O{\sc II}] or H$\alpha$,
although see Balogh \& Morris [2000]\nocite{2000MNRAS.318..703B}) to
ensure no ongoing star formation. Our selection differs primarily by
not placing limits on nebular emission, as we do not want to bias our
sample against narrow line AGN which are enhanced in the
post-starburst population \citep{2006ApJ...648..281Y,wild_psb, 2007MNRAS.382.1415S}. The additional use of
the 4000\AA\ break strength allows us to select galaxies with older
and weaker post-starburst features. 

Following the method of \citet{wild_psb} we employ a Principal
Component Analysis (PCA) of the spectra around the 4000\AA\ break to
measure the strength of the Balmer lines with sufficient SNR. In
\citet{wild_psb} the new spectral indices were defined based upon
stellar population synthesis models to avoid contamination of the
indices by gaseous emission from the galaxies.  Because of the low
spectral resolution of the VVDS spectra, in this work we are unable to
accurately mask the Balmer emission lines which fall in the center of
the Balmer absorption lines in which we are interested. Therefore, we
choose to perform the PCA directly on the VVDS spectra themselves. In
Section \ref{sec:models} we explain how we compare a posteriori to
model stellar populations created with a population synthesis code.

\subsection{Calculating the spectroscopic indices}\label{sec:indices}

\begin{figure}
\includegraphics[scale=0.7]{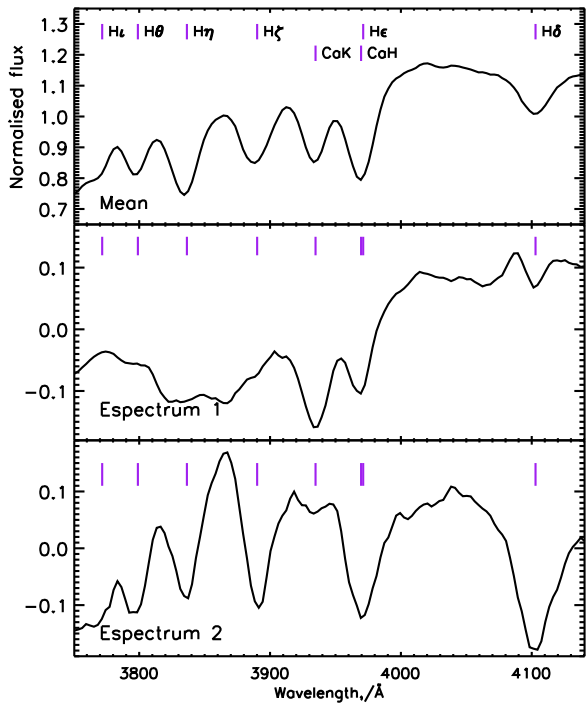}
\caption{The eigenspectra of the VVDS galaxy sample in the rest-frame
  wavelength range 3750-4140\,\AA. The spectral indices used in this
  paper refer to the amount of each eigenspectrum present in an
  individual galaxy spectrum. The first index is related to the shape
  of the spectrum and equivalent to the common D$_n$4000 index. The
  second index contains the Balmer absorption line series from
  H$\delta$ to H$\theta$, together with the distinctive decrease in
  flux bluewards of 3850\,\AA\ seen in late B to F-star
  spectra.}\label{fig:espec}
\end{figure}

All galaxy spectra in our sample are corrected for Galactic extinction
assuming a uniform E(B$-$V)$=0.027$ \citep{2003A&A...410...17M}, moved
to the galaxy rest-frame and interpolated onto a common wavelength
grid. Each spectrum is normalised, by dividing by the median flux
between 3750 and 4140\,\AA. The mean spectrum is calculated and
subtracted and a PCA is then performed on the residuals.  Because PCA
is a least-squares process, eigenspectra calculated directly from data
can easily be dominated by outliers in the input sample of spectra. In
order to obtain robust eigenspectra that are representative of the
majority of the population, we use an iterative and robust PCA
algorithm \citep{2008arXiv0809.0881B}. The algorithm essentially
replaces the least-squares minimisation solved by traditional PCA with
the minimisation of a new robust function of the data, which employs a
robust Cauchy-type function to limit the impact of outliers. The
eigenspectra are presented in Figure \ref{fig:espec}. As described in
detail in \citet{wild_psb}, the first eigenspectrum relates to the
strength of the 4000\AA\ break, and is equivalent to the well-known
index D$_n$4000. This has been shown to correlate most strongly with
the specific star formation rate of the galaxy \citep[SSFR, star
formation rate divided by stellar mass,][]{2004MNRAS.351.1151B}.  The
second index contains the Balmer absorption lines, together with the
correlated decrease in flux bluewards of 3850\,\AA\ found in late B-
to F-type stars.

We calculate the principal component amplitudes (PCs) of all galaxies
in our sample by projecting the spectra onto the eigenspectra. The
projection is performed using the ``gappy-PCA'' procedure of
\citet{1999AJ....117.2052C}, which weights pixels by their errors
during the projection and gaps in the spectra due to bad pixels are
given zero weight. We additionally allow the normalisation of the
spectra to vary as a free parameter in the projection \citep[G.~Lemson
private communication, see ][]{wild_psb}. The PCs represent the amount
of each eigenspectrum present in a galaxy spectrum and are our new
star formation history indices. Errors are calculated during the
projection of the spectra onto the eigenspectra, using the error
arrays of the individual VVDS spectra. These errors are purely
statistical, however, in Wild et~al. (2007) we show that for SDSS
spectra they compare well to the scatter observed between duplicate
observations.

\subsection{Defining the post-starburst galaxy sample}

\begin{figure*}
\includegraphics[scale=0.4]{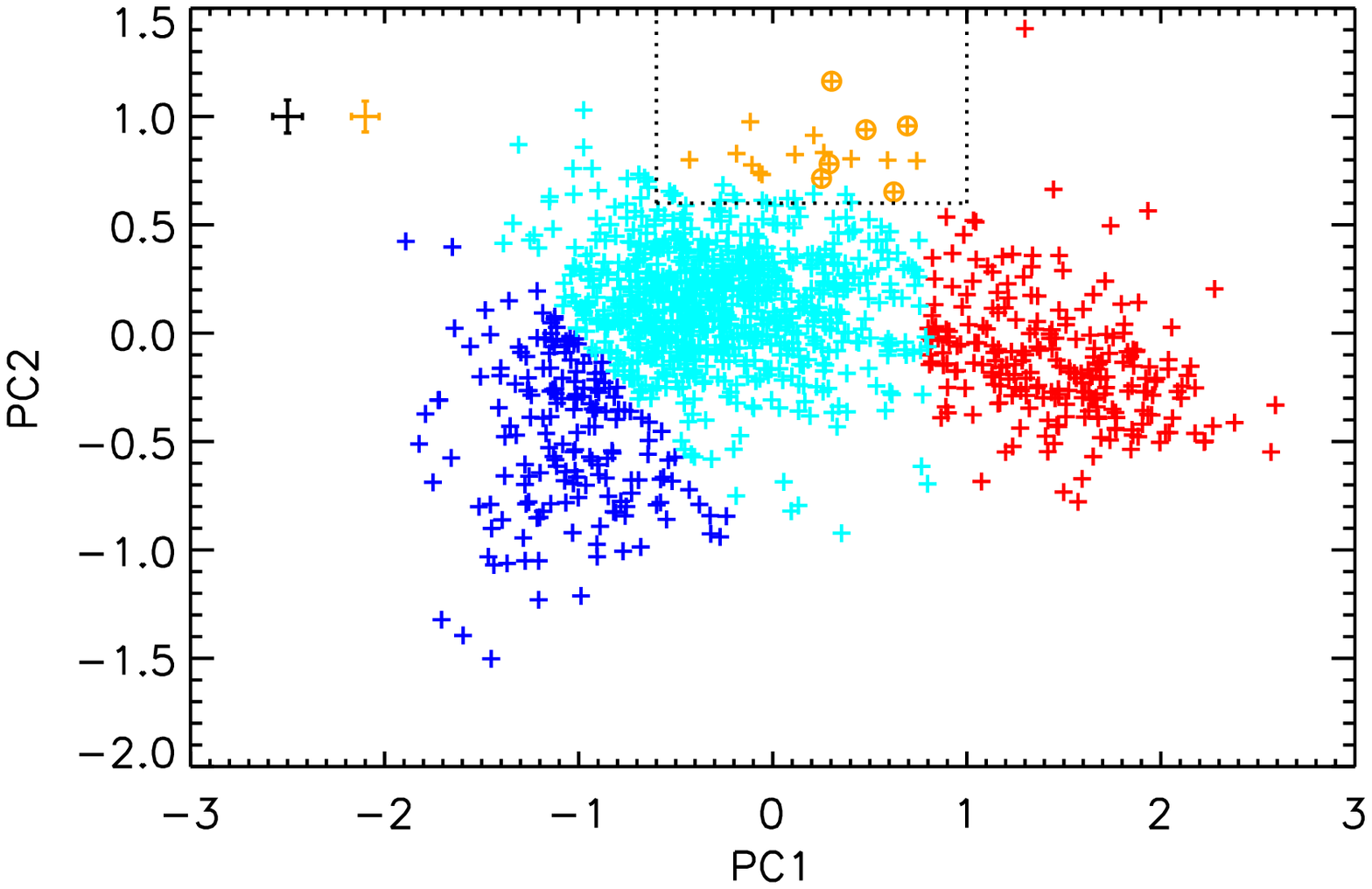}
\includegraphics[scale=0.4]{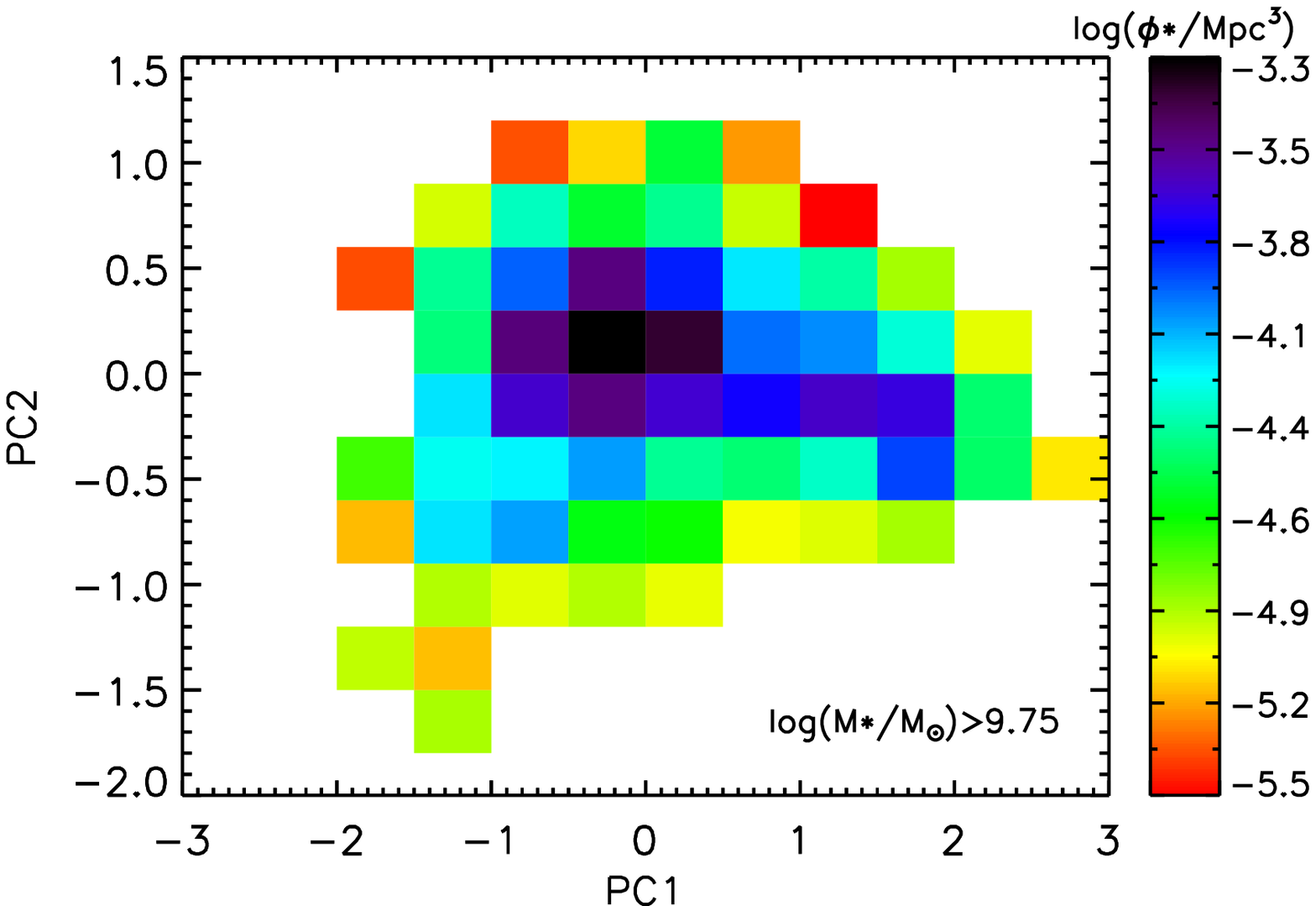}
\caption{{\it Left:} The first two principal components (PCs) for all
  galaxies in our VVDS sample. PC1 is equivalent to the well known
  index D$_n$4000, PC2 represents the excess (or lack) of Balmer
  absorption. For analysis purposes, the sample has been split into
  quiescent (red), starforming (cyan), star-bursting (blue) and
  post-starburst (orange) classes. Post-starburst galaxies are defined to lie
  within the dotted box indicated by at least $1\sigma$. Those with
  SSFR$<10^{-11}$/yr are circled. The median errors of the whole
  sample (black) and the post-starburst galaxies alone (orange) are
  shown in the top left. {\it Right:} The volume density of galaxies
  with \logm$>9.75$ in $0.5\times0.3$ bins in
  PC1/PC2. }\label{fig:pca}
\end{figure*}

\begin{figure*}
\includegraphics[scale=0.4]{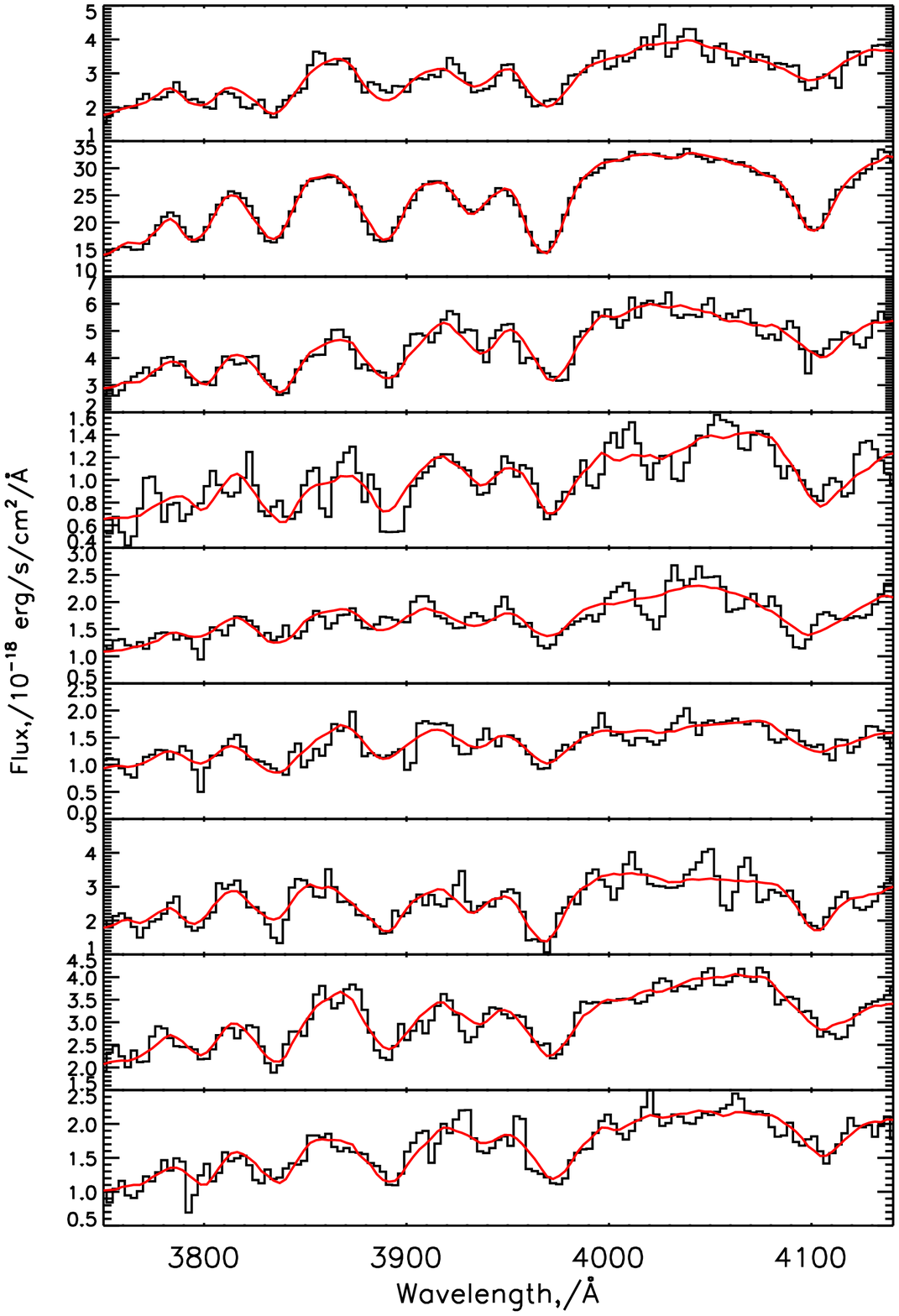}
\includegraphics[scale=0.4]{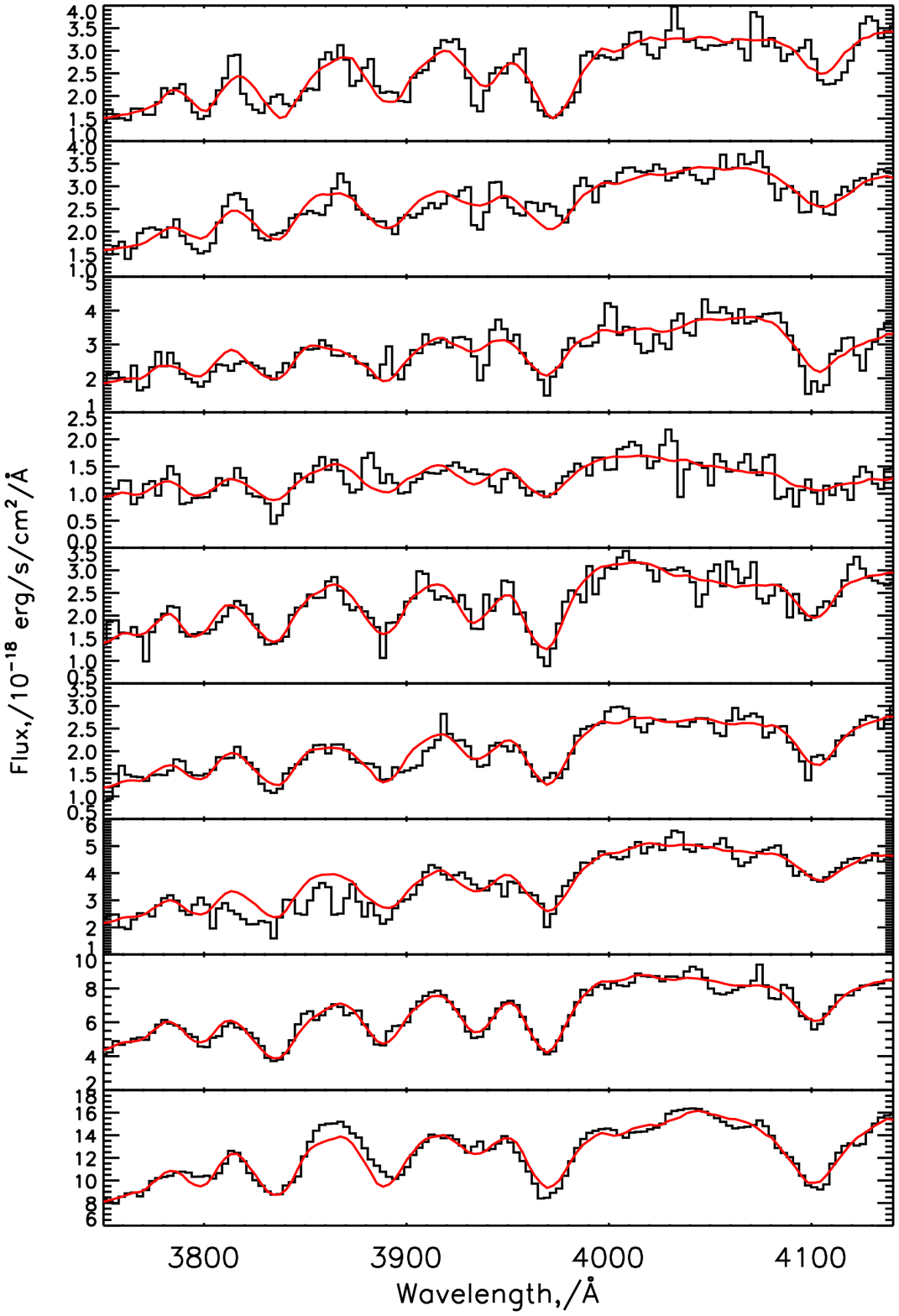}

\caption{The spectra of post-starburst galaxies culled from the VVDS
  galaxy sample in the rest wavelength range
  3750--4140\AA. Overplotted in red are the PCA fits to the
  spectra using 10 components. }\label{fig:spec}
\end{figure*}

In Figure \ref{fig:pca} we show the distribution of PC1 and PC2 for
our sample of VVDS galaxies. To guide the eye of the reader, and to
aid in further discussion, we have split the sample into four
``classes'', with boundaries empirically defined from the distribution
of points. We would like to emphasise that in reality the star
formation histories of galaxies form a continuous distribution, and
any separation into discrete classes is an over simplification,
greatly reducing the information contained within the full
distribution.  The primary division of our sample is into
``quiescent'' and ``star-forming'' galaxies on the right and left,
based on the value of PC1 (i.e. the strength of their 4000\AA\
break). The precise positioning of the boundary is arbitrary, and care
should be taken when comparing our results for ``red'' and ``blue''
galaxies to those derived from different observations.  To the bottom
left, systems with very blue continua and very strong emission lines
are found, which we class as ``starburst'' galaxies.

Finally, to the top center we find the ``post-starburst'' galaxies,
with stronger Balmer absorption lines than expected for their 4000\AA\
break strength. We define ``post-starburst'' galaxies to have
$-0.6<{\rm PC1}<1.0$ and PC2$>0.6$. While these boundaries were
defined empirically from the distribution of data (this class are
clear outliers in PC2), we will later compare with model stellar
populations to justify their classification as ``post-starburst''. We
find that 18 galaxies lie in this region at $>1\sigma$ confidence.
The spectra of these galaxies are shown in Figure \ref{fig:spec}. The
selection of these 18 galaxies is robust to changes in the exact
method used to create the eigenspectra.

\subsection{Estimating completeness limits}

It is important to estimate the completeness of the survey in terms of
galaxy stellar mass. This of course depends upon the mass-to-light
ratio and SED shape of the galaxies in question: flux limited surveys
are generally complete to lower mass limits for star forming galaxies
than for quiescent galaxies.  While the completeness in a particular
photometric band can be judged empirically, this is not possible for
the completeness in derived parameters, such as stellar mass and star
formation rate. We therefore use an alternative approach to estimate
the mass completeness of our sample, described in detail in W08.

Because we are primarily interested in the post-starburst population,
with mass-to-light ratios between those of star-forming and quiescent
galaxies, we derive the mass limits relevant for this population
alone. From the model dataset described in Section \ref{sec:Ms} we
select all model galaxies which have undergone a starburst in the last
Gyr and lie within our PSB region in PC1/PC2. For a given stellar
mass, we determine the fraction of model post-starburst galaxies that
lie within our survey magnitude limits. We find that 50\% of the model
post-starburst galaxies at $z=0.6$ with \logm$>9.5$ would be detected,
and \logm$>9.9$ at $z=0.9$. The stochastic model library assumed in
this analysis has a higher fraction of strong post-starburst galaxies
than appear in the real data, i.e. while the observational PC1/PC2
space is fully covered by the models, the relative space density of
the different galaxy types is not reliable. We therefore confirmed
that the mass completeness limits did not vary greatly with burst mass
fraction or burst age.

In the following sections, all results will be quoted for a mass limit
of \logm$>9.75$. We have confirmed that increasing this limit to
\logm$>10.0$ does not alter our conclusions. Figure \ref{fig:pca}
(right) shows the number density of galaxies in bins of PC1/2 for
galaxies with \logm$>9.75$, corrected for survey volume and sampling
effects.


\section{Comparison with models}\label{sec:models}

In order to understand the particular properties of a starburst which
would lead to a post-starburst galaxy in our sample, and to measure
the timescale during which a post-starburst galaxy is visible, it is
necessary to compare our data with model stellar populations. We begin
by creating simple toy model starbursts and subsequently turn to the
analysis of smoothed particle hydrodynamic (SPH) simulations of
merging galaxies.

\subsection{Toy model galaxy tracks}
 
\begin{figure*}
\includegraphics[scale=0.4]{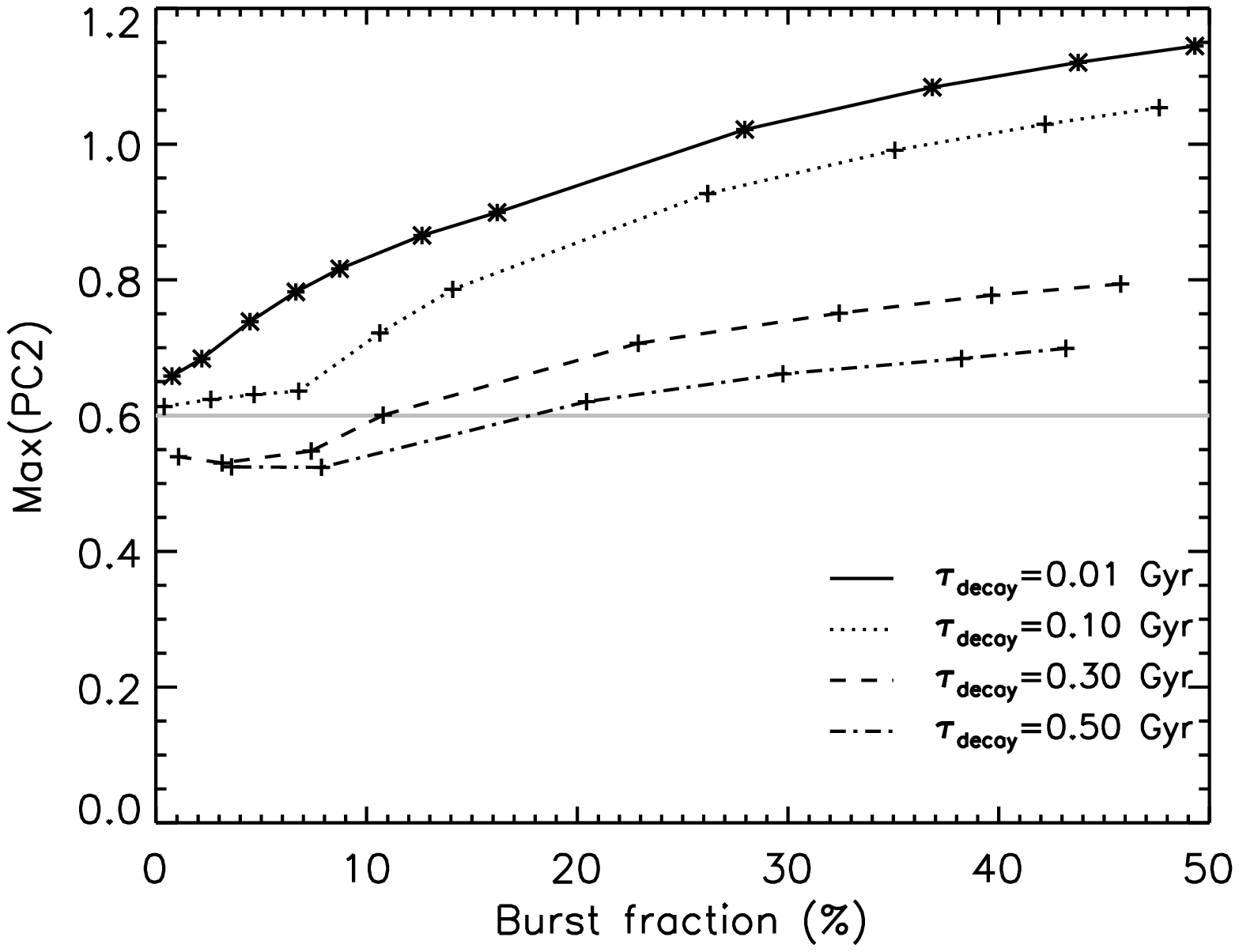}
\includegraphics[scale=0.4]{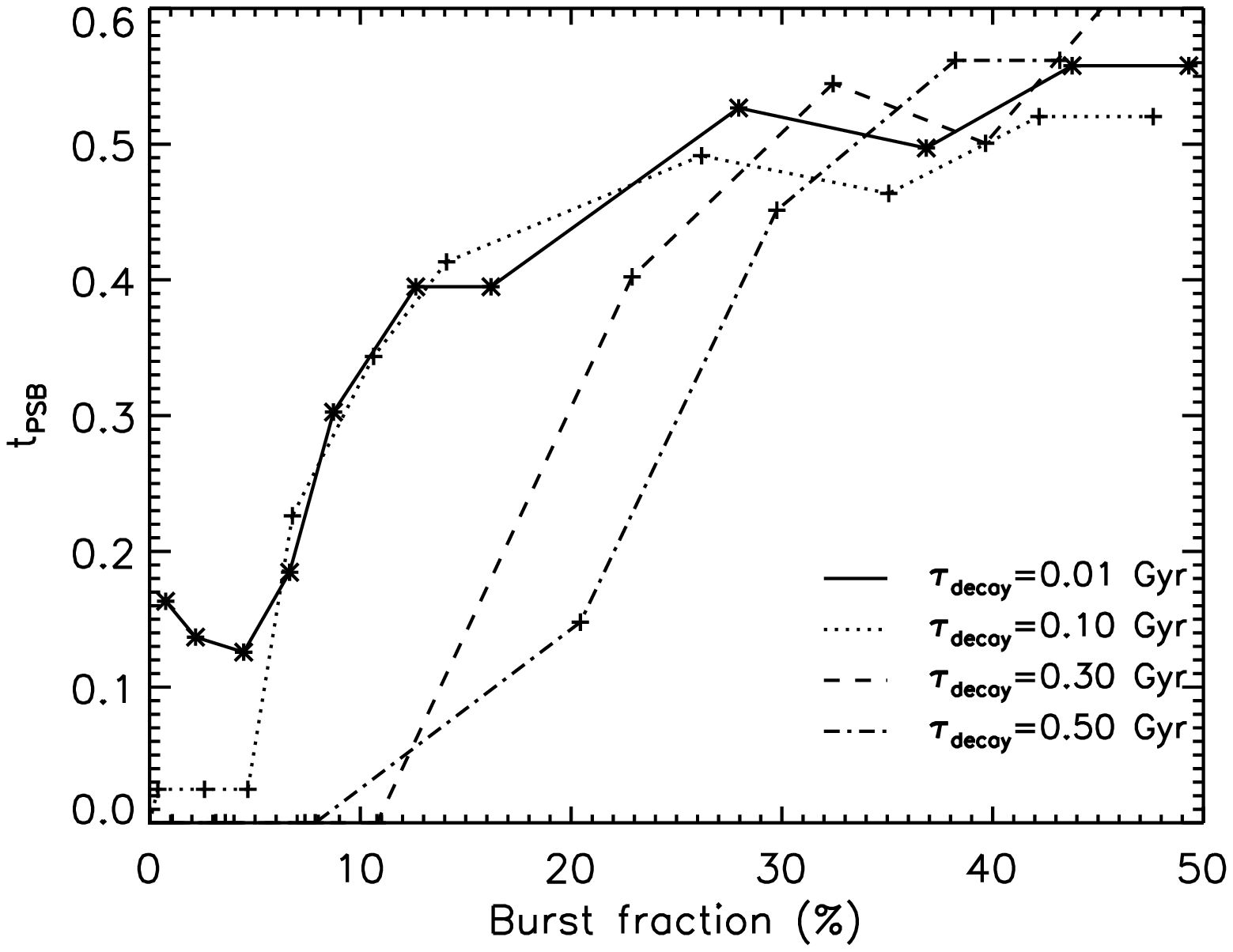}
\caption{By projecting evolving model stellar populations onto the
  VVDS eigenspectra, we can measure quantitative properties relevant
  to our PSB galaxies. After continuously forming stars for 5\,Gyr,
  the toy-model galaxies undergo a starburst which declines
  exponentially with decay time $\tau_{decay}$.  {\it Left:} The
  maximum PC2 attained by the models after the starburst has occurred,
  i.e. a measure of the strength of the PSB features. Different lines
  represent models with bursts of different $\tau_{decay}$. {\it
  Right:} The time spent in the PSB phase as a function of burst mass
  fraction.}\label{fig:mock}
\end{figure*}

For a galaxy to reveal strong Balmer absorption lines, star formation
must switch off rapidly resulting in an excess of longer lived A- and
F-stars over the hotter O- and early B-stars. This does not
necessarily have to be preceded by a short-lived starburst, the
sudden truncation of ordinary star formation will suffice, although
the signature will of course be weaker. The two parameters of
importance for the strength and longevity of the post-starburst phase
are the mass fraction of stars formed in a starburst and the decay
timescale of the starburst. In this terminology, a truncation model
then simply has a mass fraction of zero per cent.

With this in mind we have created a library of model galaxies which
undergo exponentially decaying bursts, occurring after 5\,Gyr of
continuous star formation. These bursts are parameterised by their
decay time and burst mass fraction, i.e. the mass of stars formed
while the star formation rate is above the continuum level.

Converting the star formation histories into spectra suitable for
comparison with the VVDS dataset requires several steps. We input the
star formation histories into the {\sc GALAXEV} code
\citep{2003MNRAS.344.1000B}, using Charlot \& Bruzual (in preparation)
simple-stellar-populations (SSPs), a \citet{2003PASP..115..763C} IMF
and assuming solar metallicity. The evolution of the model galaxies in
PC1/2 is not greatly affected by the change between
\citet{2003MNRAS.344.1000B} and Charlot \& Bruzual (in preparation)
SSPs, which involves both a change in stellar libraries and in the
stellar evolution tracks. Solar metallicity was chosen as being
suitable for our post-starburst galaxy sample, given their stellar
masses (Section \ref{sec:physprop}), but the evolutionary tracks are
insensitive to moderate variations in metallicity. The two component
dust prescription of \citet{2000ApJ...539..718C} is applied to the
continuum light as implemented in the {\sc GALAXEV} code. Again, the
inclusion or not of continuum dust has little effect on the time spent
on the PSB sequence, or the maximum PC2 observed. This is
because the dominant A/F star population in the post-starburst spectra
is assumed to already have emerged from the dense stellar birth
clouds.

Next, because the resolution of the VVDS spectra is too low to mask
emission lines in the data, we must add emission lines to the model
spectra. We convert the rate of ionising photons predicted by the
stellar population synthesis model ($Q_{ion}$) into H$\alpha$ line flux
using:
\begin{equation}
F_{H\alpha} [erg/s] =  0.45 * E_{H\alpha}[erg] * Q_{ion}[s^{-1}]
\end{equation}
where E$_{H\alpha}$ is the energy of a H$\alpha$ photon and 0.45 is
the fraction of ionisations which lead to the emission of an H$\alpha$
recombination photon \citep[case B, ${\rm Te}=1\times 10^4$K and ${\rm
Ne}=1\times 10^4$cm$^{-3}$,][]{1995MNRAS.272...41S}. The remaining
Balmer line fluxes are derived from the ratios given in
\citet{1989agna.book.....O}. The stellar population synthesis
model includes the post-AGB phase of stellar evolution which is
believed to be important for producing ionising photons in older
stellar populations \citep{1994A&A...292...13B}. We note that no
contribution to the emission lines from obscured, narrow-line AGN is
included in the models. Any substantial infilling of the Balmer
absorption lines in this way will prevent us from identifying
post-starburst galaxies. AGN with strong enough lines to have a
significant effect are extremely rare however, and their absence from
our sample will not affect our final results.

Just as dust attenuation is applied to the model stellar continua, the
lines are similarly attenuated using the attenuation law as described
in \citet{wild_psb}:
\begin{equation}
\frac{\tau_\lambda}{\tau_V} =(1-\mu)(\frac{\lambda}{5500{\rm
  \AA}})^{-1.3} + \mu (\frac{\lambda}{5500{\rm \AA}})^{-0.7}.
\end{equation}
This law is inspired by the two-component dust model of
\citet{2000ApJ...539..718C}. For the birth clouds, from which
nebular emission is believed to originate and which contain a
fraction $1-\mu$  of the dust, these authors adopted for
simplicity an attenuation law to match that of the diffuse
interstellar medium (i.e. of the form $\lambda^{-0.7}$). Our slightly steeper curve,
resulting in slightly increased attenuation, is motivated by the fact
that dense clouds around young stars have a more shell-like geometry
than the diffuse interstellar medium \citep[see][for more
details]{2008MNRAS.388.1595D}.  The attenuation of the model emission
lines by a moderate amount ($A_V=1.0$) alters slightly the
evolutionary tracks in PC1/PC2 of the model during the starburst
phase.

Our ability to detect a post-starburst galaxy depends on the maximum
PC2 reached during the evolution of the stellar population, and the
time spent in the post-starburst phase ($t_{\rm PSB}$). These factors
depend in turn on the strength of the proceeding starburst, together
with the timescale of the subsequent decay of the star formation rate
($\tau_{\rm decay}$). Figure \ref{fig:pca} shows that the maximum
PC2 reached by the VVDS galaxies is around 1.0, with 5/18 having PC2
$>$0.9. Using our toy models, the left panel of Figure
\ref{fig:mock} shows the maximum value of PC2 attained by a model galaxy
after the burst has occurred, as a function of the burst mass and burst
decay time. The models show that the descendants of starbursts with
decay times $\la0.1$\,Gyr and burst fractions above a few percent will
appear clearly in the post-starburst phase. The right panel of Figure
\ref{fig:mock} shows the time spent in the post-starburst phase
($t_{\rm PSB}$). This time is short, $\la$0.15\,Gyr, for burst mass
fractions $<5$\%, and reaches a maximum independent of decay time of
$\sim$0.6\,Gyr for mass fractions of $\ga$30\%. We will use this
value as an upper limit on $t_{\rm PSB}$ in later sections.

We conclude from this section that the majority of the progenitors of
our PSB galaxies underwent a recent strong starburst, followed by a
rapid truncation of the star formation. A small fraction may however
be simply the result of rapid truncation of their ongoing star
formation.

\subsection{Simulation galaxy tracks}\label{sec:mergers}
The next step in our comparison with models involves a sample of 
79 merger simulations presented in \citet[][hereafter
JNB09]{2008arXiv0802.0210J}. 

\subsubsection{Simulation details}

The simulations were performed using the entropy conserving
TreeSPH-code GADGET-2 \citep{2005MNRAS.364.1105S}, which includes
radiative cooling for a primordial mixture of hydrogen and helium
together with a spatially uniform time-independent local UV
background.  Star formation and the associated supernova feedback is
implemented using the sub-resolution multiphase model developed by
\citet{2003MNRAS.339..289S}.  We model the feedback from black holes
following the effective subresolution model of
\citet{2005MNRAS.361..776S}, in which the unresolved accretion onto
the black hole (BH) is related to the resolved gas distribution using
a Bondi-Hoyle-Lyttleton parameterization \citep{1944MNRAS.104..273B}.
Further details concerning the feedback implementations and parameter
choices can be found in JNB09. In addition, for the purposes of this
paper a new suite of simulations was performed in which the BH
feedback was switched off. Such a comparison is important because the
BH feedback can cause star formation to completely shut down after the
merger, which may plausibly affect both the time the merger remnant
spends in the PSB phase, and the strength of the post-starburst
features.

The simulations include mergers of gas-rich disks (Sp-Sp), of
early-type galaxies and disks (E-Sp, mixed mergers), and mergers of
early-type galaxies (E-E, dry mergers).  The progenitor galaxies have
a range of virial velocities, $v_{\rm vir}$ of 80, 160, 320 and
500\,km/s, which determines their masses and sizes (see Eqs. 1 \& 2 in
JNB09). All model galaxies contain an exponential disk ($d$)
component, together with a stellar bulge ($b$) embedded in a dark
matter halo modelled with the \citet{1990ApJ...356..359H} profile. The
stellar bulge has a fixed mass ratio of $M^*_{\rm b} = 0.01367 M_{\rm
vir} = 1/3 M_{\rm d,tot}$, where the total disk mass is a sum of the
disk stellar and gas mass $M_{\rm d,tot}=M^*_{\rm d}+M_{\rm d,gas} =
0.041 M_{\rm vir}$, the disk stellar mass is $M^*_{\rm d} = 0.041
(1-f_{\rm gas}) M_{\rm vir}$, and $M_{\rm vir}$ is the virial mass of
the galaxy. The initial gas fraction $f_{\rm gas}$ takes values of
0.2, 0.4 and 0.8.  The galaxies approach each other on parabolic
orbits, which are generally motivated by statistics from N-body
simulations \citep{2006A&A...445..403K}.  The resulting merger
remnants have total stellar masses between \logm$=9.5-12.5$. Both
major (1:1) and minor (3:1) mergers were simulated with varying
orbital and initial disk geometries \citep{2003ApJ...597..893N}. All
simulations were evolved for a total of $t=3 \ \rm Gyr$, with the
merger taking place after approximately 1.5\,Gyr, using the local
Altix 3700 Bx2 machine hosted at the University Observatory in
Munich. In this paper we will primarily concentrate on mergers with
prograde in-plane orbits \citep[``G0'', i.e. the galaxies collide
directly edge on][]{2003ApJ...597..893N}. We will additionally show an
example in which galaxies rotate in a retrograde sense, with the disks
tilted by 30 degrees with respect to one another (``G7'').  We will
focus exclusively on the Sp-Sp mergers, as all other pair combinations
were found not to produce the strength of PSB galaxy visible to us in
the data.

As the simulations, and thus star formation rates, are only computed
over a total duration of 3\,Gyr, we must assume a star formation
history and galaxy age for the initial build up of stellar mass in
each progenitor galaxy.  We assume that by the start of the simulation
each galaxy has built both its bulge and disk stellar mass through an
exponentially declining star formation rate: ${\rm SFR} = {\rm SFR}_0
+ n \exp(-t/\tau)$ where $\tau=1{\rm Gyr}$. SFR$_0$ is the initial SFR
of the simulation i.e. $\sim$1.5\,Gyr before the merger takes
place. $n$ is defined such that the galaxies are 5\,Gyr old at the
start of the simulation, i.e. the stellar populations are
approximately 7.5\,Gyr old at the onset of the merger. Both
progenitors in a simulation are assumed to undergo the same initial
SFH. We note that any similar star formation history can be assumed
with which to build the pre-merger galaxies, without significantly
affecting their evolution in the post-merger post-starburst phase.
The creation of the spectra from the SFHs of the simulations follows
exactly the same method as for the toy burst models.

\subsubsection{Post-starburst galaxies in merger simulations}

\begin{figure*}
\includegraphics[scale=0.5]{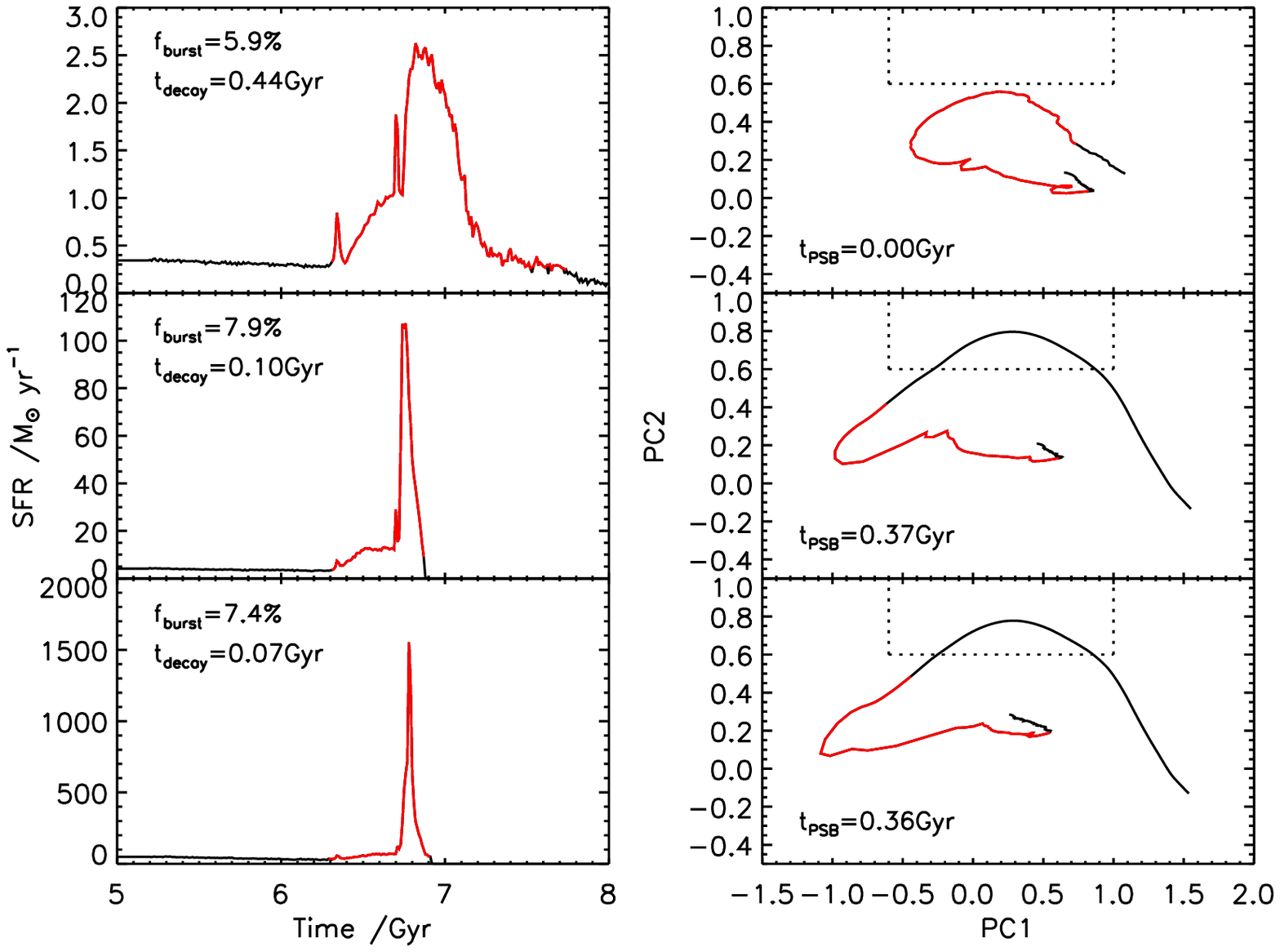}
\includegraphics[scale=0.5]{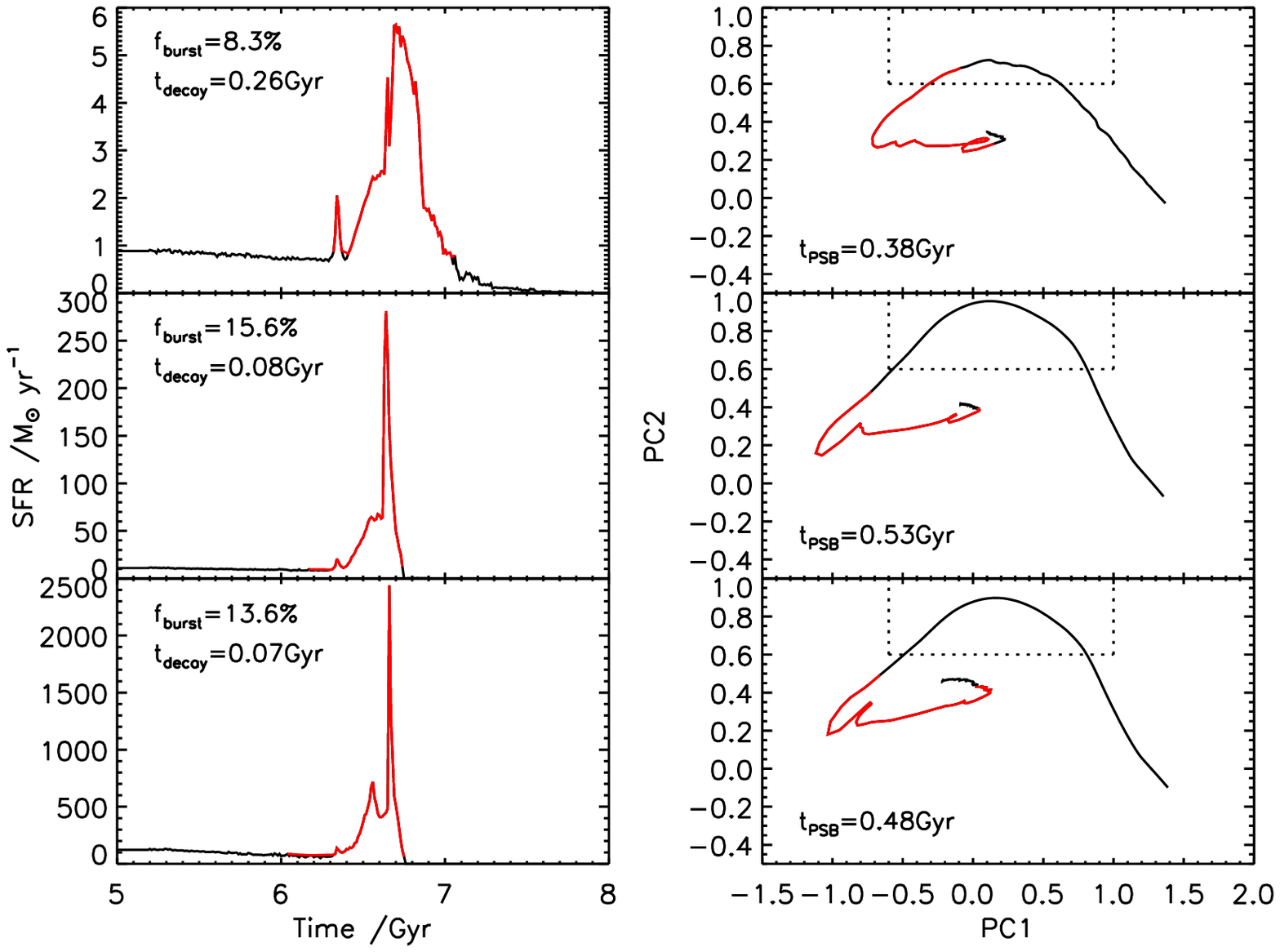}
\caption{ {\it Far left:} The combined star formation history of both
  galaxies participating in a major merger during the 3\,Gyr SPH
  simulation. In these panels, the galaxies collide in a ``G0''
  orbital configuration and BH feedback is included. Progenitor
  galaxies have gas fractions of 20\% and virial velocities of 80
  (top), 160 (middle) and 320 (lower)\,km/s corresponding to stellar
  masses of around $10^{10}$, $10^{11}$ and $10^{12}$M$_\odot$. The
  red part of the SFH depicts the ``starburst'' phase, defined to be
  where the SFR is 5\% greater than the continuum SFR. The values in
  the top left of the panels are the mass fraction of stars formed
  during the starburst and the decay timescale of the starburst (see
  text). {\it Next left:} the corresponding evolutionary tracks in
  PC1/PC2, with the red section corresponding to the red section in
  the left panel. The dotted box shows the PSB region defined in this
  paper. The time that the merged galaxy spends in the PSB phase is
  given in the bottom left of the panels. {\it Right pair of panels:} The same as the
  left panels, but for progenitor galaxies with gas fractions of
  40\%. }\label{fig:simn1}
\end{figure*}

\begin{figure*}
\includegraphics[scale=0.5]{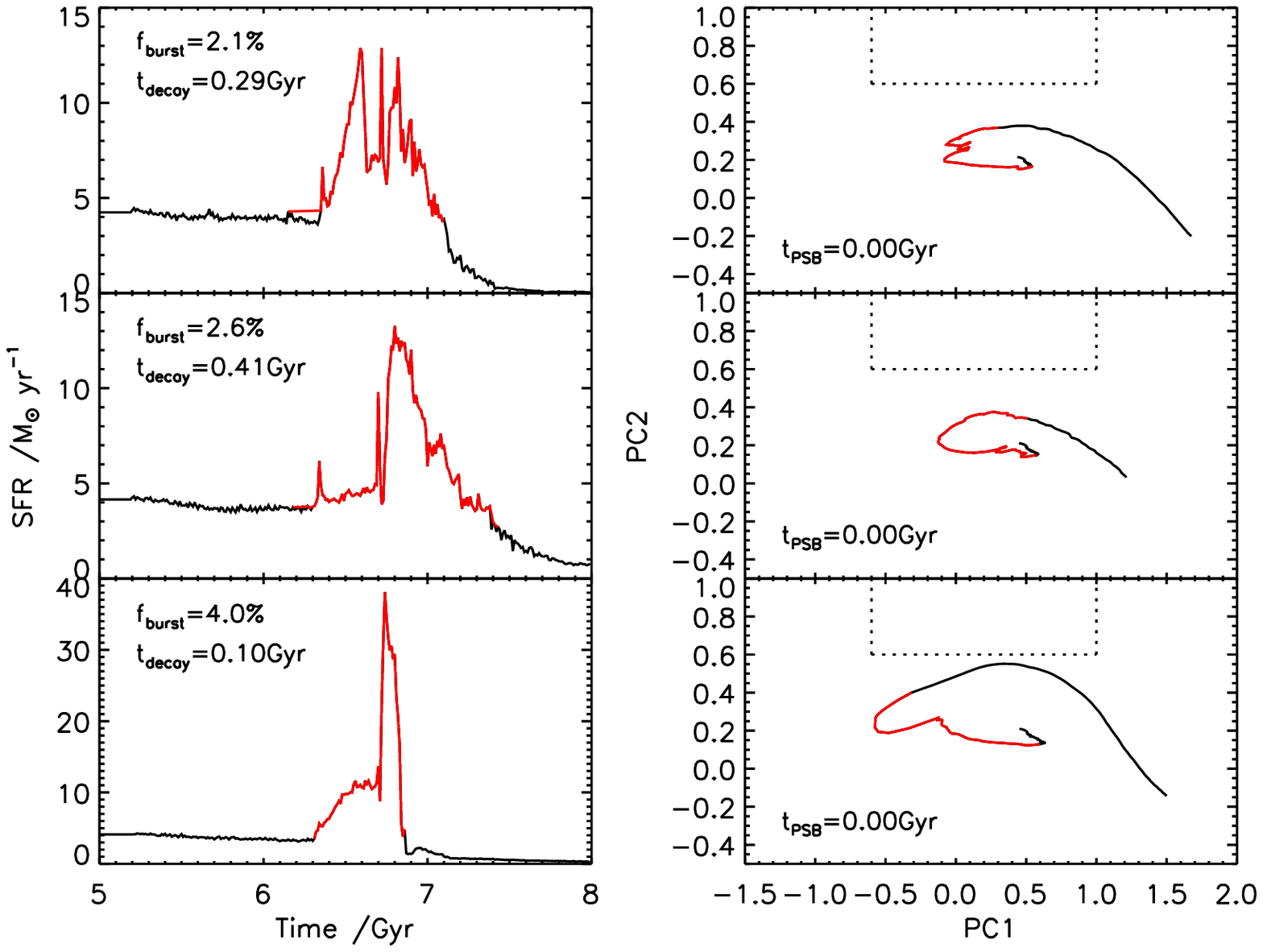}
\includegraphics[scale=0.5]{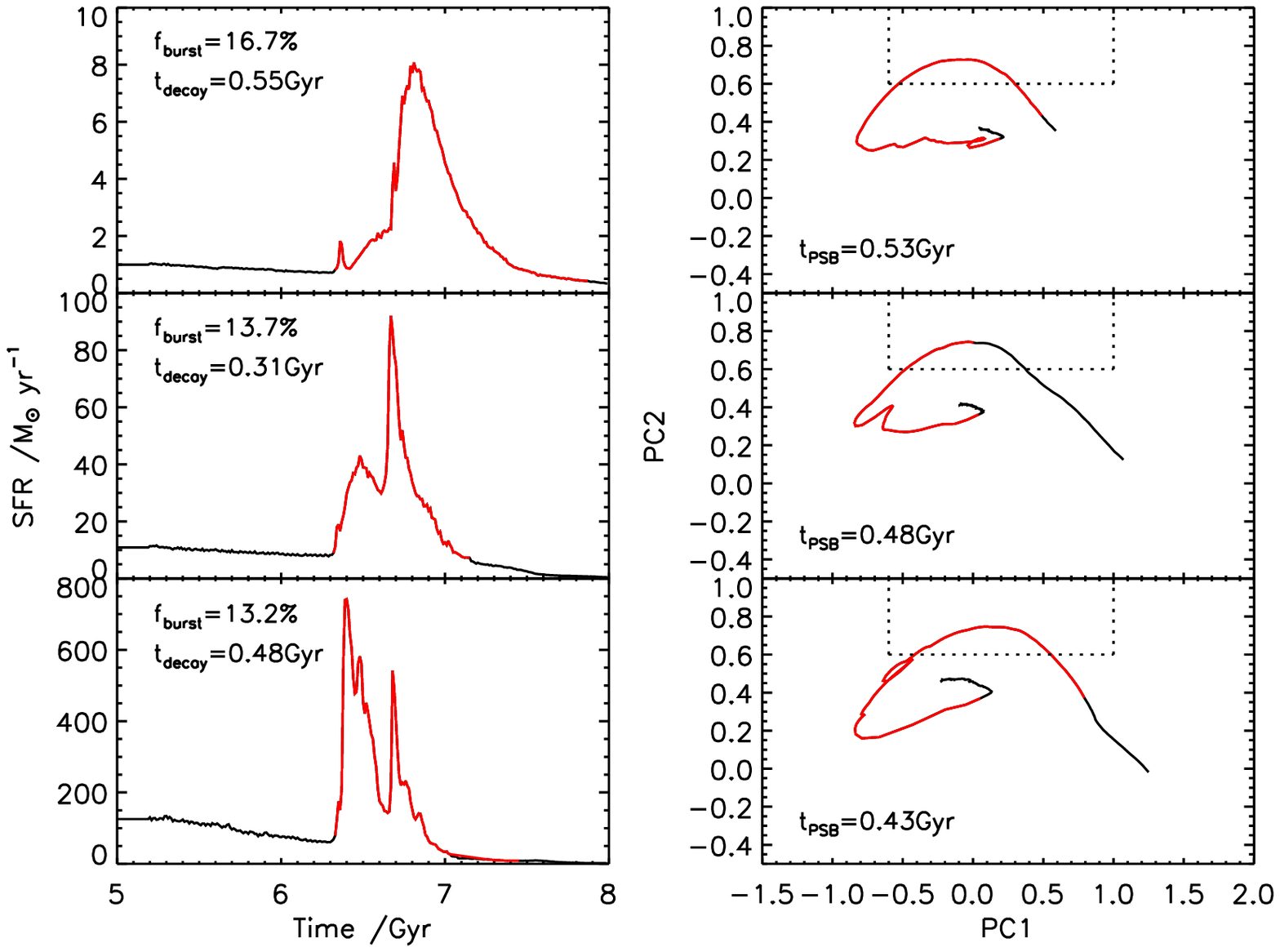}
\caption{Same as for Figure \ref{fig:simn1} but for different initial
 conditions. {\it Left pair of panels:} Simulations in which the
 progenitor galaxies collide with inclinations other than
 edge-on. From top to bottom the orbits are ``G7'', ``G10'' and
 ``G13''. The galaxies have gas mass fractions of 20\% (compare to
 left hand panels of Figure \ref{fig:simn1}. {\it Right pair of
 panels:} Simulations in which the black hole feedback has been
 switched off. The progenitor galaxies have gas mass fractions of 40\%
 (compare to right hand panels of Figure \ref{fig:simn1}).}\label{fig:simn2}
\end{figure*}

\begin{figure*}
\includegraphics[scale=0.4]{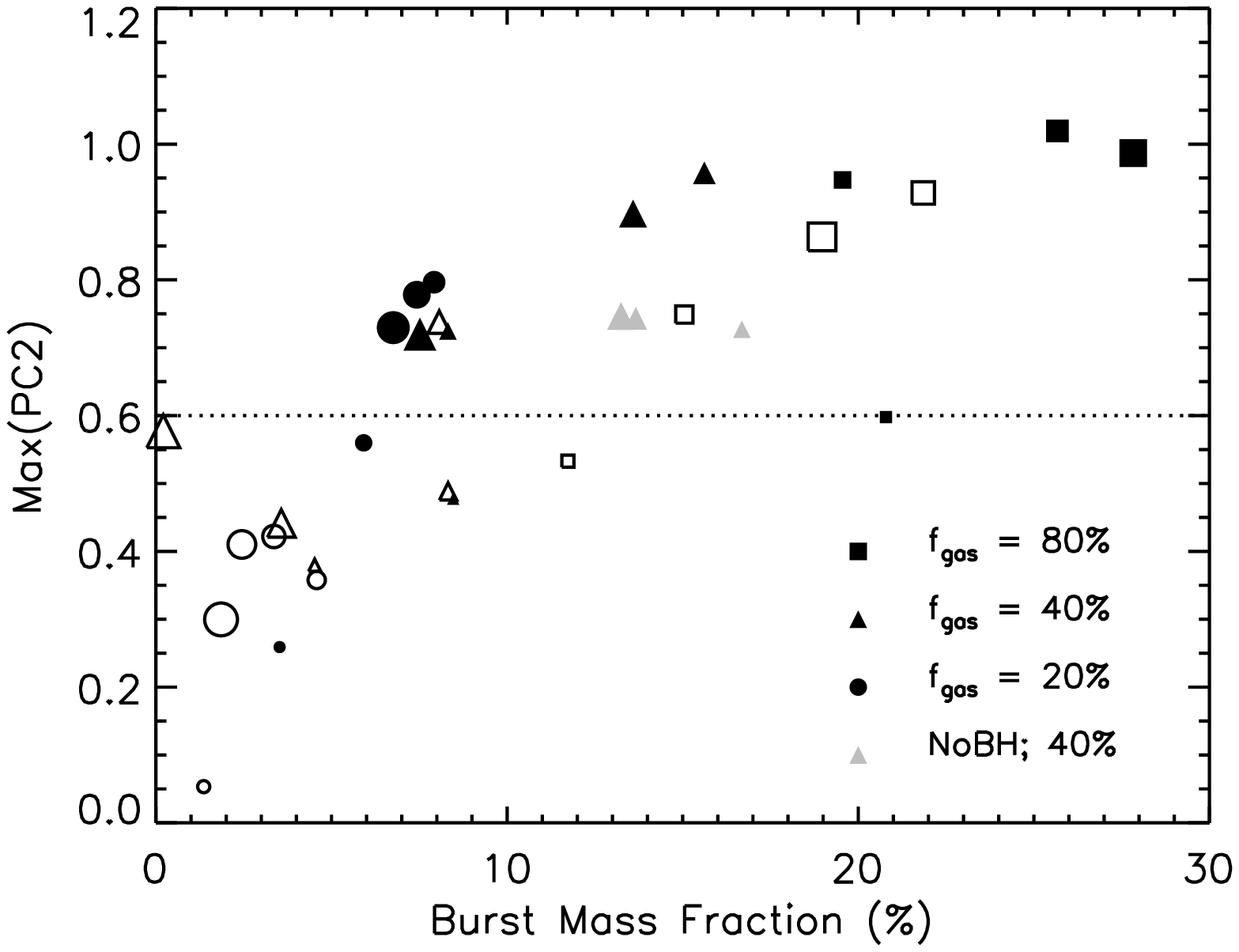}
\includegraphics[scale=0.4]{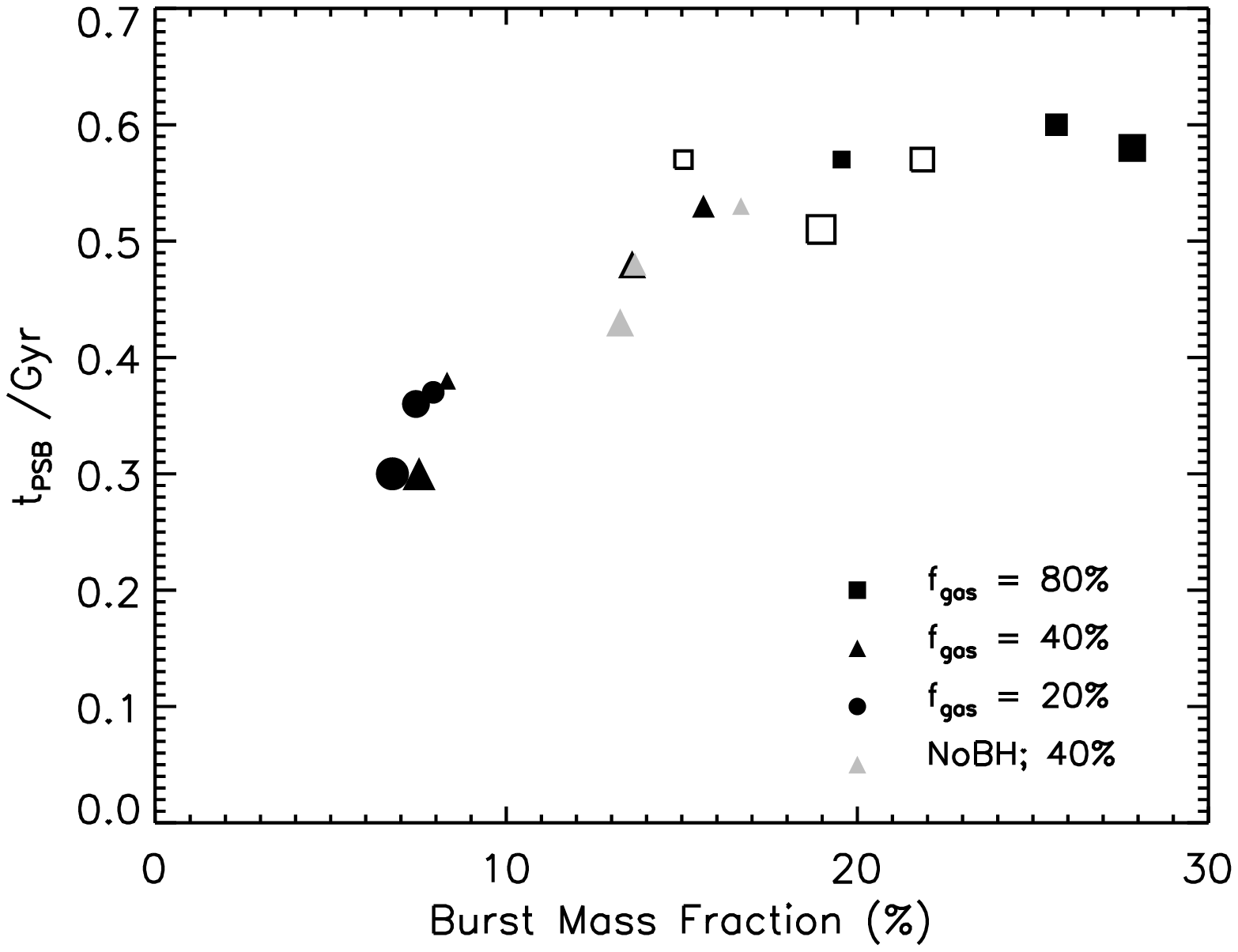}
\caption{ Same as Figure \ref{fig:mock} except for SPH galaxy merger
  simulations with ``G0'' orbits. Different shapes represent different
  gas fractions of the progenitor galaxies, filled symbols are 1:1
  mass ratio mergers and open symbols are 3:1 mass ratio
  mergers. Increasing symbol size denotes increasing stellar
  mass. Gray symbols are mergers with black hole feedback ``switched-off'',
  40\% gas fractions and $v_{\rm vir}=80$, 160 and 320\,km/s.  {\it
  Left:} The maximum PC2 attained by the models after the starburst
  has occurred, as a function of burst mass fraction. The dotted line
  indicates our observational limit above which PSBs are
  detected. {\it Right:} The time spent in the PSB
  phase.}\label{fig:peter}
\end{figure*}

Figure \ref{fig:simn1} shows the star formation history and
evolutionary tracks in PC1/2 for three of the major merger simulations
with $f_{gas}=0.2$ (left) and $0.4$ (right), BH feedback and prograde,
edge-on orbits (``G0'') over the 3\,Gyr of the simulation. From top to
bottom panel the mass of the progenitor galaxies increases. The SFHs
of the major merger simulations in general follow similar patterns,
with an initial weak enhancement in their star formation, quickly
followed by the starburst and decay to zero star formation. 

We parameterise the simulations using two simple values to describe
the evolution of the starburst. Firstly, the burst mass fraction
$f_{\rm burst}$, defined as the mass of stars formed while the SFR is
at least 1.05 times the continuum rate, divided by the total mass of
the final merger remnant. The continuum rate is measured in the first
Gyr of the simulation, and linearly extrapolated across the subsequent
starburst. The second parameter is the decay time of the starburst,
$t_{\rm decay}$, defined to be the time from the peak of the starburst
until the SFR is only 10\% of the maximum burst height. These values
are shown in the top left of each SFH panel in Figure \ref{fig:simn1}. As
found using the toy-models, higher gas fractions lead to stronger
post-starburst features and longer periods of time spent in the
post-starburst phase.

Using this suite of simulations we can investigate several other
initial conditions which affect the evolution of the stellar
population of the merger remnant: progenitor mass ratio; orbital
geometry; presence or absence of a black hole. We illustrate the
latter two in Figure \ref{fig:simn2}. The left two panels show the
star formation history and evolutionary tracks in PC1/2 for galaxies
with tilted orbital configurations and $f_{gas}=0.2$. Comparing to the
left panels of Figure \ref{fig:simn1} we see that in this particular
suite of simulations, only the prograde edge-on mergers give rise to
post-starburst features. While the starbursts are weaker in the other
orbital configurations, it is likely that the main cause of the
difference is in the longer decay times of the star formation rate.
However, we caution that the tilted orbital geometries tested here
have typical tilts of $\ga30$ degrees and are retrograde. Other
relevant parameters, such as impact parameter, galactic structure and
variations in stellar feedback prescription
have not been investigated. While the simulations indicate that a
relatively strong interaction is required to produce a post-starburst
galaxy \citep[see also][JNB09]{2008MNRAS.384..386C}, a
full study of the post-merger stellar populations as a function of
impact geometries is beyond the scope of this paper.

The right two panels of Figure \ref{fig:simn2} show the evolution of the
stellar population with BH feedback ``switched-off'' and
$f_{gas}=0.4$. Comparing to the right-hand panels of Figure
\ref{fig:simn1} we see that the burst mass fractions and time spent in
the post-starburst phase are similar with or without the presence of
BH feedback. The primary causes of the shut-down in star formation are
supernovae feedback and exhaustion of gas supplies, rather than any
additional energy supplied by the AGN.  In two out of three cases the
post-starburst features are stronger when BH feedback aids in
expelling the gas, but the main effect is focused on the later stages
of the burst where any residual star formation is halted more
quickly. Therefore, we conclude that from these spectral indices
alone, there is no way to confirm or refute the suggestion that strong
BH feedback provides the mechanism for shutting off star formation in
the Universe.

In Figure \ref{fig:peter} we summarise the results by showing the time
spent in the post-starburst phase and maximum value of PC2 reached
after the merger has occurred, for all of the Sp-Sp merger simulations
with prograde in-plane orbits (``G0'', i.e. directly edge on).  The
results paint a very similar picture to that shown by the toy-models,
but we can now associate burst mass fractions to physically meaningful
parameters such as gas fractions and galaxy masses. We find that
mergers must induce burst mass fractions of at least 5\% for them to
be detectable in our observationally defined PSB phase, and the time
spent there is greater for more gas rich and more massive mergers. For
a 3:1 merger (open symbols) to appear in the PSB phase, a very high
gas fraction of 80\% is apparently required. The duration of the
post-starburst phase tends to a maximum of 0.6\,Gyr, with lower gas
fractions leading to shorter timescales of 0.3-0.4\,Gyr and burst mass
fractions of 5-10\%. Similar to the toy models, we find that decay
times of the starburst must be shorter than $\sim$0.3\,Gyr for a
post-starburst galaxy to be observed. The results of three simulations
with BH feedback ``switched-off'', $f_{gas}=0.4$ and a range of
virial masses, are shown as grey symbols. These can be compared
directly to the filled triangles in the same figure, in which BH
feedback has occurred. We note that the burst mass fractions and
starburst decay timescales derived in this section are the same as
those found for luminous infra-red galaxies
\citep[LIRGS, ][]{2006A&A...458..369M}.

Many of the simulated galaxy mergers in the full library of 79
simulations do not result in a post-starburst remnant according to our
demanding observational criteria. Even for gas rich major mergers, the
starbursts can occasionally be weak and/or have long decay times, thus
not creating the sharp discontinuity in SFR required for them to
appear in the post-starburst phase. Such galaxies may enter the red
sequence through the ``green valley'' if their star formation rate
continues to decay.


\section{Mass densities and mass fluxes of post-starburst galaxies at
  z$\sim$0.7}  

In the preceding sections we have culled a sample of post-starburst
galaxies with $0.5<z<1.0$ from the VVDS redshift survey. Through
comparisons with toy models and SPH galaxy merger simulations, we have
shown how the spectroscopic features indicate that the galaxies have
undergone a recent major, gas rich merger. In this section, we will
place the post-starburst galaxies into a global context, in terms of
their stellar masses, star formation rates and their number and mass
densities. Finally, we will address the interesting question: how much
mass enters the red sequence through the post-starburst phase?

\subsection{Photometrically derived physical properties}\label{sec:physprop}

\begin{figure*}
\includegraphics[scale=0.5]{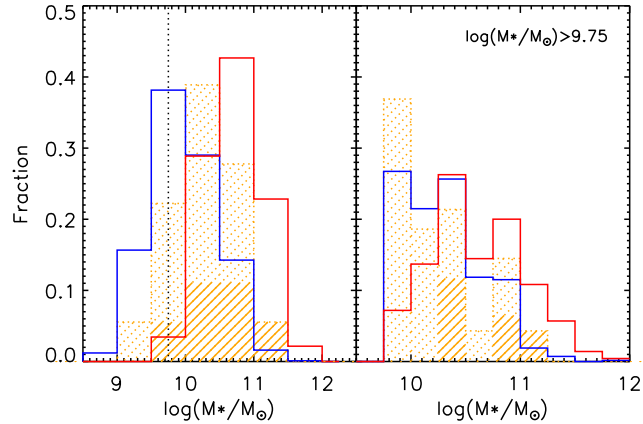}
\caption{The mass distribution of galaxies, classified into
  starforming (blue), quiescent (red) and post-starburst (orange,
  dot-filled) by their spectral type i.e. their PC1/PC2 values shown
  in Figure \ref{fig:pca}. The orange line-filled histograms indicate
  the subset of post-starburst galaxies with no ongoing residual star
  formation.  The verticle dotted line indicates our post-starburst
  mass completeness limit. {\it Left:} the raw data, with no
  correction for survey incompleteness effects. {\it Right:} for
  galaxies with stellar masses above our completeness limit, after
  correction for survey incompleteness effects. }\label{fig:mass}
\end{figure*}

We can now ask how the mass of the post-starburst galaxies compares to
that of the blue- and red-sequence galaxies? Figure \ref{fig:mass}
shows the mass distributions of our different classes of galaxies,
before and after corrections for sample selection (TSR, QSR and SSR)
and survey volume. The galaxies have been classified according to
their spectral types i.e. position in PC1/PC2. For clarity, we have
combined the starburst and starforming galaxies into a single class;
their mass distributions are very similar. The left hand panel shows
the mass distribution before correcting for completeness effects. The
right hand panel shows the completeness corrected mass distribution,
indicating that the post-starburst galaxies primarily have
\logm$<10.5$, i.e. to the lower end of galaxies in the red sequence
and with a similar distribution to galaxies on the blue sequence.
It is also worth noting that the completeness corrected mass
distribution of the post-starburst galaxies continues to rise to lower
masses, and does not turn over before we reach the completeness limit
of the sample. Deeper spectroscopic surveys will be required to fully
probe the mass distribution of post-starburst galaxies.

\begin{figure*}
\includegraphics[scale=0.4]{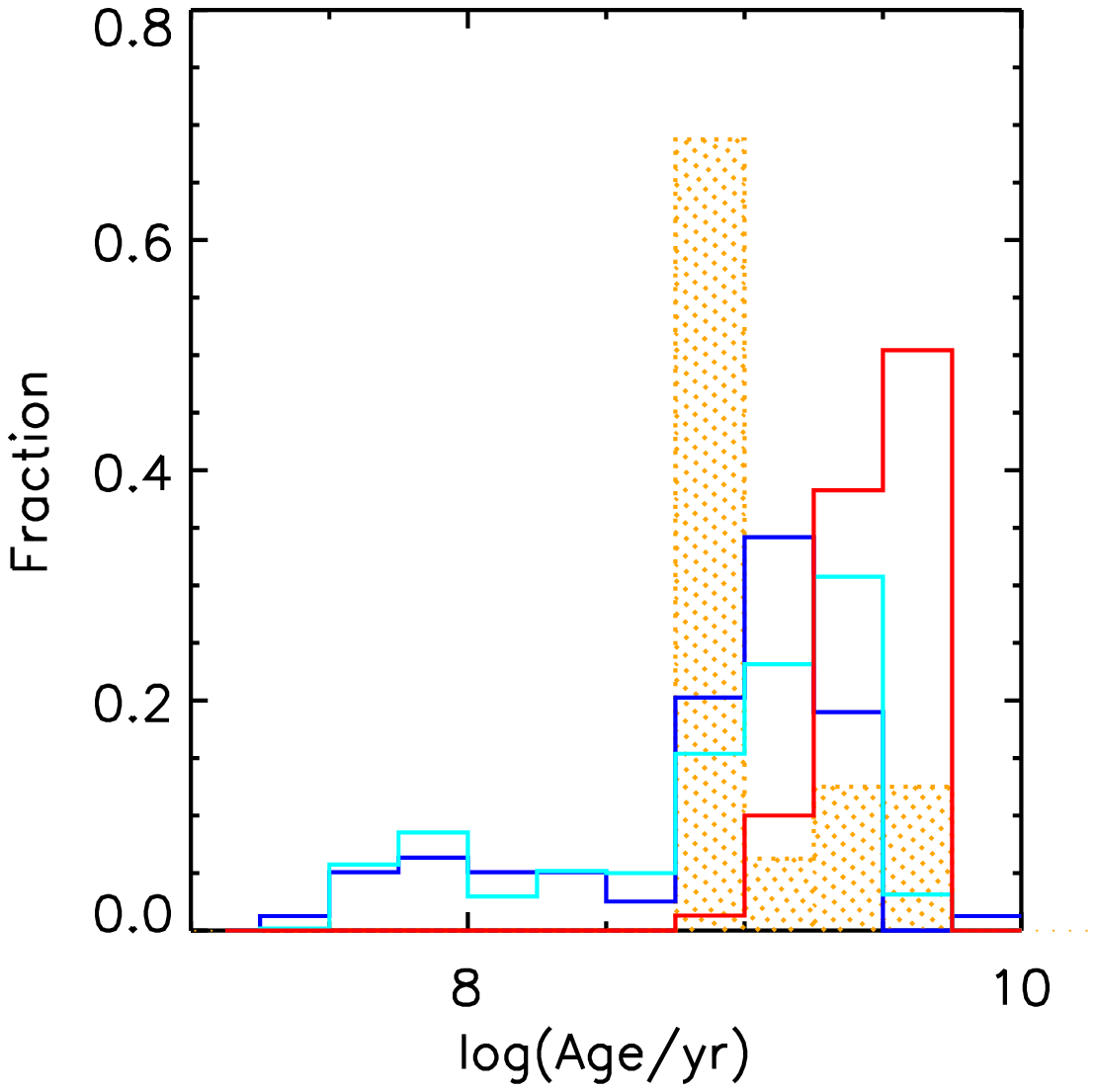}
\includegraphics[scale=0.4]{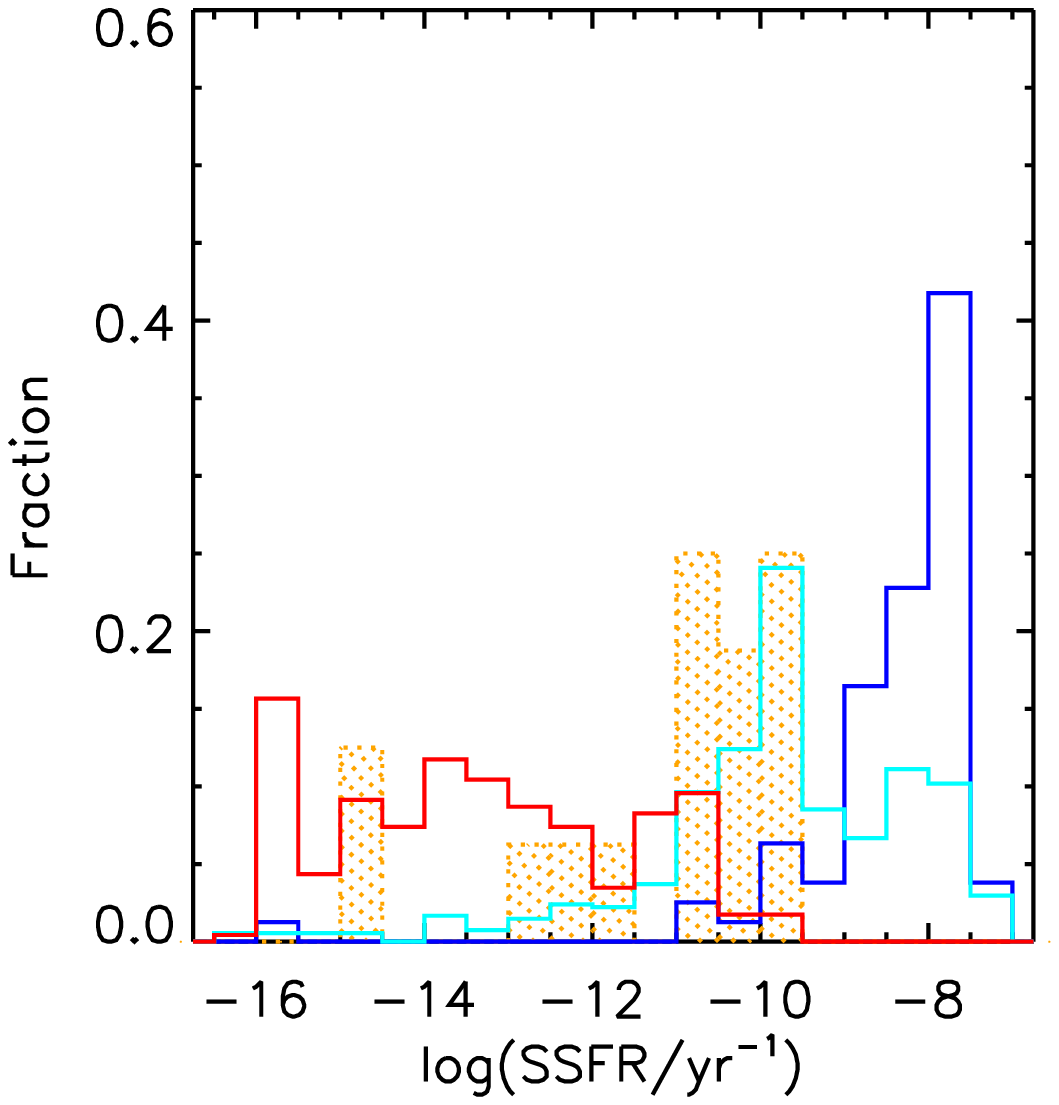}\\
\includegraphics[scale=0.4]{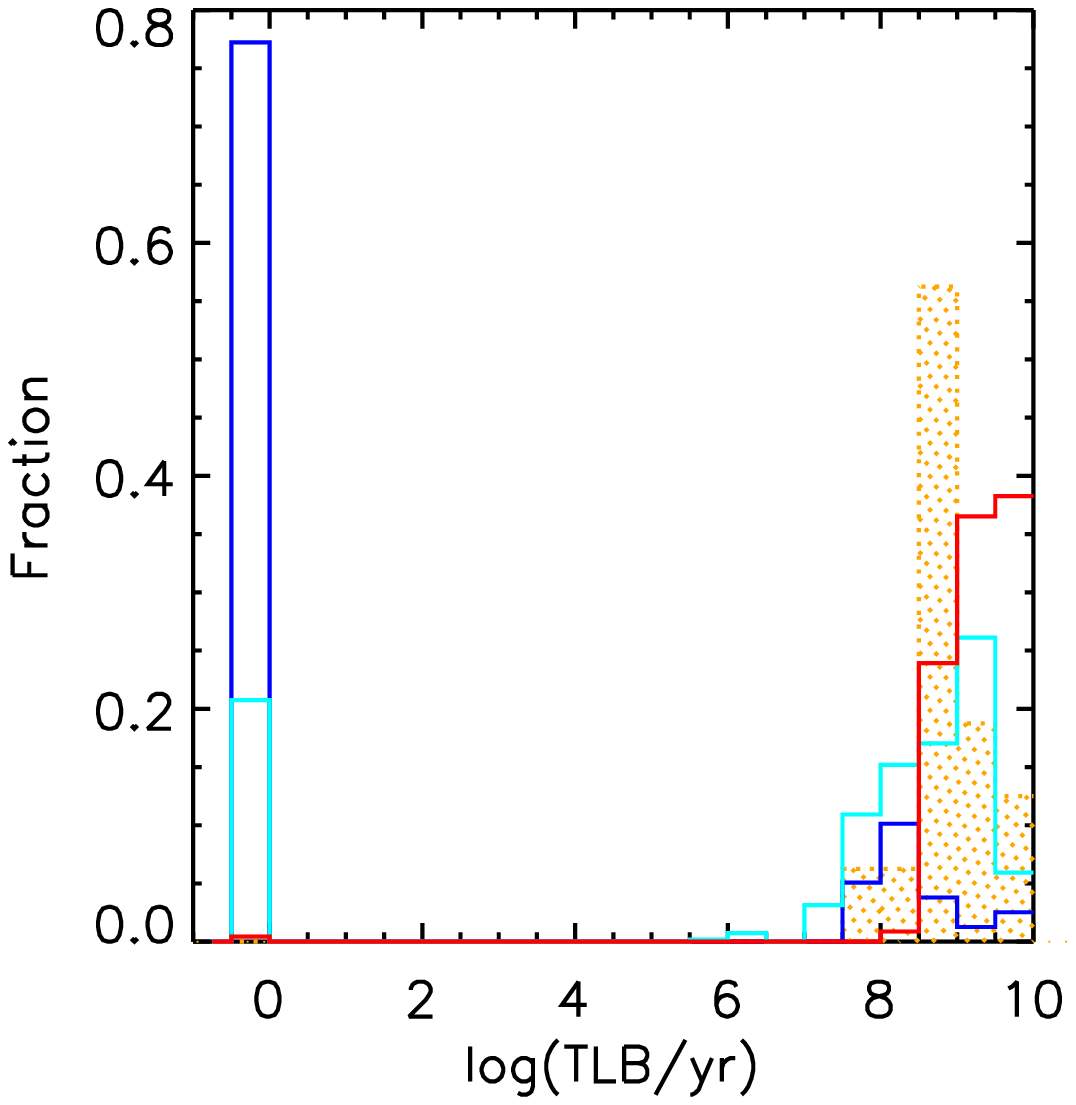}
\includegraphics[scale=0.4]{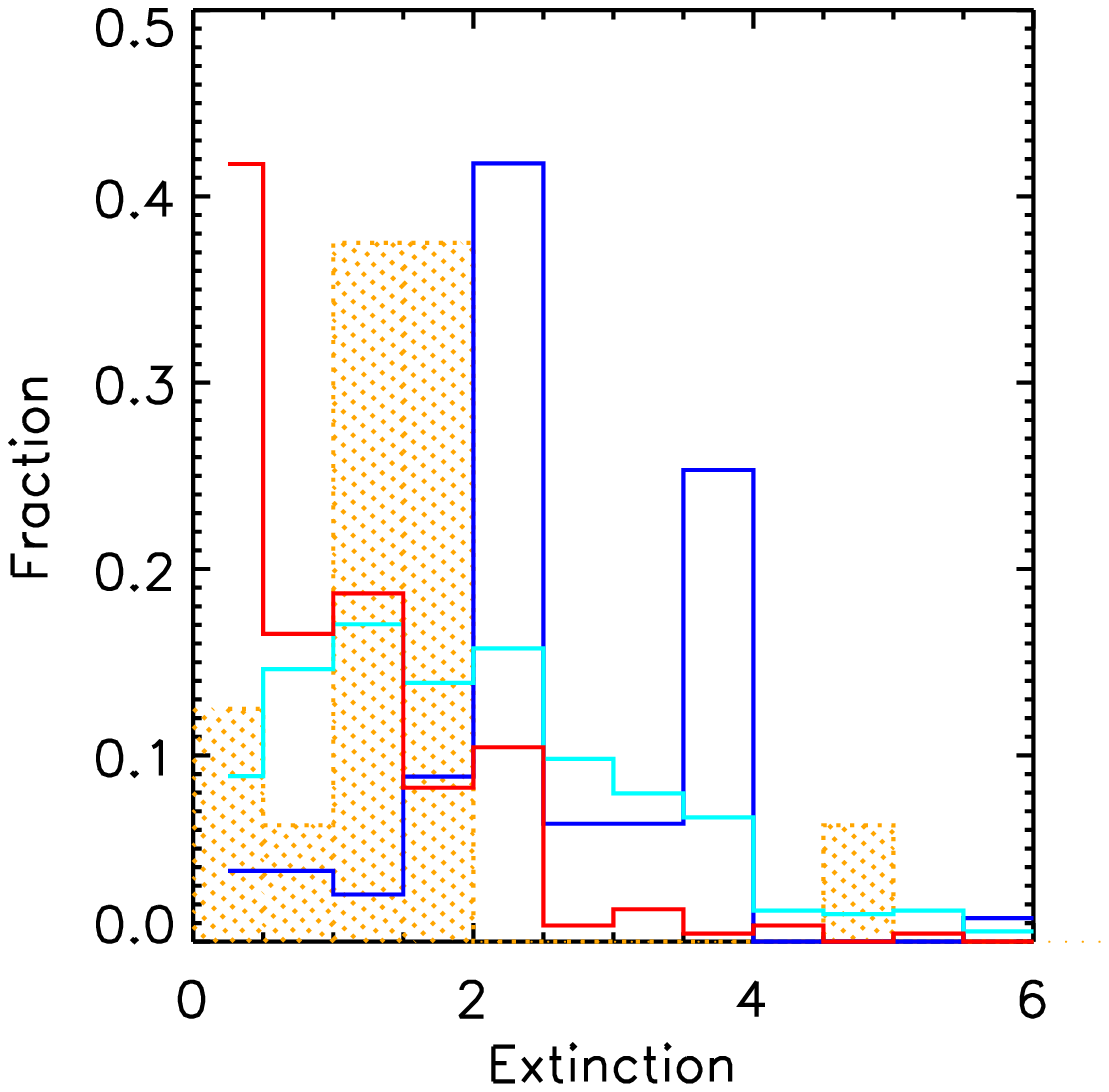}
\caption{The physical properties of our spectroscopic galaxy sample
  above a mass limit of $\log(M/M_\odot)>9.75$, derived from fitting their
  multiwavelength SEDs \citep[see][]{2008arXiv0807.4636W}. From top left,
  clockwise: the light-weighted mean stellar age; the specific star
  formation rate (SFR/M$^*$); the time of the last starburst; the
  extinction in the $V$ band. The galaxies
  have been classified by their PC1/2 values into quiescent (red),
  starforming (cyan), starburst (blue) and post-starburst
  (orange). The histograms represent the raw data, they are not
  corrected for incompleteness effects.}\label{fig:physprop}
\end{figure*}

Our spectroscopic analysis has focused on uncovering the special
class of post-starburst galaxies, and not on recovering further
physical parameters. As described in Section \ref{sec:Ms} and more
fully in W08, the multiwavelength (UV--IR) SEDs of these galaxies have
been analysed to obtain more physical properties than just stellar
mass. In Figure \ref{fig:physprop} we present the light-weighted mean
stellar age, specific star formation rate (SFR/M$^*$), time of last
burst and dust content of our full sample of galaxies, classified by
their PC1/PC2 values. Overall, the relative agreement between the SED
fitting and VVDS spectral fitting is remarkable: galaxies which are
spectroscopically classified as starforming are found, from
multi-wavelength photometry, to be younger, have high star formation
rates, more recent starbursts and higher dust contents compared to
galaxies spectroscopically classified as quiescent. The
multi-wavelength photometry is able to distinguish that the
post-starburst galaxies have ages and star formation rates
intermediate between the quiescent and star-forming
galaxies. Additionally, the estimated time of their last starburst has
a lower mean value than quiescent galaxies, peaking at $10^8$--$10^9$
years, exactly as expected from the VVDS spectral analysis. We can see
that the post-starburst galaxies, identified purely through their
stellar continuum as having undergone a recent starburst, have a range
of specific star formation rates which place them at the lower end of
the star-forming sequence through to the quiescent galaxies. In Section
\ref{sec:massflux} we will use this result to set a lower limit on the
mass flux onto the red sequence through the PSB phase, by identifying
those galaxies which have completely ceased star formation.

\subsection{Mass and number density}\label{sec:massdens}

\begin{table*}
  \begin{center}
  \caption{The mass and number densities of each of our classes of
  galaxies, for a mass limit of \logm$>9.75$. \label{tab:mass}}

\vspace{0.2cm}

  \begin{tabular}{cccccc} \hline\hline
  Class  & Quiescent & Star-forming & Starburst  & Post-starburst & Total \\ \hline
  $\log({\rm Number/Mpc}^3)$ & -2.5 & -2.9 & -3.1& -4.0& -2.3\\
  $\log(\rho^*/{\rm M_\odot/Mpc}^3)$ & $7.89^{+0.01}_{-0.01}$ &
  $7.88^{+0.02}_{-0.02}$ & $7.17_{-0.04}^{+0.03}$ & $6.43_{-0.04}^{+0.06}$ &
  $8.23_{-0.01}^{+0.01}$\\   
  \hline
  \end{tabular}\\
  \end{center}
\end{table*}

Summing up all the galaxies in our sample above the mass completeness
limit of $\log M=9.75$ with the appropriate weightings, we obtain a
total stellar mass density of $\log\rho^* =
8.23_{-0.011}^{+0.013}$. The errors account for individual errors on
the stellar masses and Poisson errors from the SSR and QSR weights. As
we have placed a stringent criterion on the minimum signal-to-noise
ratio of the spectra, it is important to ensure that we recover the
correct total stellar mass density found in other
work. \citet{2007A&A...474..443P} calculate the total mass density of
galaxies as a function of redshift in the VVDS spectroscopic sample,
using the same \citet{2003PASP..115..763C} IMF as used in the present
work. Integrating a Schechter function with the parameters\footnote{We
compare to the $I$-band selected sample with masses determined from
complex star formation histories (row 7 of their Table 2), for better
comparison with our own sample and masses} given by Pozzetti
et~al. for the redshift range $0.7<z<0.9$ above our mass limit of 9.75
gives a stellar mass density of $\log\rho^* = 8.28$.  Although the
statistical errors are always $<10\%$, Pozzetti et al. quote a typical
scatter between different methods of mass determination and luminosity
function estimates of at least $\sim0.1$dex. Taken together with the
slightly different redshift range, our results appear to be entirely
consistent with those of Pozzetti et al..

Table \ref{tab:mass} gives the mass and number densities for each of
our galaxy classes. Note that the completeness of our survey is lower
for quiescent galaxies than for other classes.

Focussing on the post-starburst galaxies, we find a number density of
of $1.0\times10^{-4}$ per Mpc$^3$ and mass density of
$2.69\times10^6$M$_\odot$/Mpc$^3$. It is instructive to compare their
number density to other objects considered to be unusual in the
Universe at these redshifts. The number density of our post-starburst
sample is 0.3-0.5dex above the number density of powerful X-ray
selected AGN with 2-8\,KeV luminosities above $10^{43}$erg/s at
redshifts $0.5<z<1.5$ \citep{2005AJ....129..578B}. They are are factor
of 10 more numerous than star-forming galaxies at these redshifts with
radio luminosities similar to local ultra luminous infra-red galaxies
\citep[ULIRGS,][]{2004ApJ...603L..69C}. The post-starburst galaxies
are considerably more common than sub-mm selected galaxies with FIR
luminosities $\ga6\times10^{11}$L$_\odot$ at $z\sim0.9$, which have
number densities of $3\times10^{-6}$/Mpc$^3$/decade
\citep{2005ApJ...622..772C}. These comparisons are designed for
orientation only, and not to imply evolutionary sequences: the very
different duty cycles and ``on'' times for different classes of rare
objects make inferences about the relation between them difficult. We
will return to a quantitative comparison with QSO number densities and
major merger rates in Section \ref{sec:disc}.

\subsection{The mass flux of post-starburst galaxies onto the red sequence}\label{sec:massflux}

We are interested in how much mass may enter the red sequence after a
starburst and subsequent fast quenching of the star formation as seen
in simulations of major mergers. We estimate the total mass flux which
passes through the post-starburst phase to be:
\begin{equation}\label{eq:massflux}
\dot{\rho}_{A \rightarrow Q, PSB} =
  \frac{1}{t_{PSB}}\sum_{i=1}^{N_{PSB}} \frac{M^*_i}{w_i . 
  V_{max,i} }
\end{equation}
where $M^*$ is the stellar mass of the galaxy in solar masses, $w$ is
the combination of the selection functions $w^{\rm SSR}$, $w^{\rm
TSR}$ and $w^{\rm QSR}$, and $V_{max}$ is the correction for the
volume in which each galaxy can be seen. The sum is over
post-starburst galaxies above the mass completeness limit of
\logm$>9.75$. $t_{PSB}$ is the time that a galaxy will be seen in the
post-starburst phase. From Section \ref{sec:models} we find $t_{PSB}$
to be less than 0.6\,Gyr. Galaxy mergers with lower gas fractions of
around 20-40\% result in slightly smaller burst mass fractions and
shorter time spent in the post-starburst phase of $\sim0.35$\,Gyr.

During the selection of our post-starburst galaxy sample, we have
imposed no restrictions on nebular emission line strengths (see Section
\ref{sec:dead}).  However, galaxies which show post-starburst
signatures as well as nebular emission lines may be in the process of
regenerating their star formation and may therefore subsequently return to the
blue-sequence. In order to derive a firm lower limit on the mass flux
entering the red sequence through the PSB phase, we elect to sum only
those post-starburst galaxies which show no evidence of star
formation. Inspecting the distribution of SSFRs of our post-starburst
galaxies (top right panel of Figure \ref{fig:physprop}) shows the
majority have SSFRs similar to the lower end of the star-forming
sequence, 6 objects have SSFRs below $10^{-11}$ per yr, 5 of which lie
above our mass completeness limit. We note that these galaxies also
have no measurable [O{\sc II}] emission. The mass distribution of
these galaxies is indicated in Figure \ref{fig:mass} by the
filled-line histogram and their positions marked in PC1/2 by the open
circles in Figure \ref{fig:pca}.

Summing Equation \ref{eq:massflux} over the 5 post-starburst galaxies
above the mass completeness limit with negligible ongoing star
formation as derived from their SEDs, assuming an upper limit for
$t_{PSB}$ of 0.6\,Gyr, results in a lower limit on the mass flux
entering the red sequence through the post-starburst phase of
$\dot{\rho}_{A \rightarrow Q, PSB} > 0.0022^{+0.0002}_{-0.006}$
M$_\odot$/Mpc$^3$/yr.  Our best estimate for the mass flux through the
post-starburst phase onto the red sequence, assuming
$t_{PSB}=0.35$Gyr, is $\dot{\rho}_{A \rightarrow Q, PSB} =
0.0038^{+0.0004}_{-0.001}$ M$_\odot$/Mpc$^3$/yr. We note that this is
still a lower limit, as we have excluded all those post-starburst
galaxies with detectable levels of ongoing star formation. This
residual star formation may still decay as the post-starburst ages, we
note that the younger post-starburst galaxies (those with smaller PC1)
are found to be more likely to have residual star formation than older
post-starburst galaxies (see circled points in Figure
\ref{fig:pca}). If all the post-starburst galaxies were to enter the
red sequence, for $t_{PSB}=0.35$Gyr the mass flux would be
$\dot{\rho}_{A \rightarrow Q, PSB} = 0.0077$ M$_\odot$/Mpc$^3$/yr

To determine how important this mass flux is in terms of the global
build up of the red sequence, we compare to
\citet{2007A&A...476..137A}. They use the multiband photometric data
in the same VVDS-02h field to measure the mass build up of the red
sequence; comparison with this result thus limits our exposure to
cosmic variance. The galaxy masses are measured using the same code
and model dataset as in this paper, thus eliminating one more
potential source of systematic error. They find\footnote{They quote a
value of $\dot{\rho}_{A \rightarrow Q} = 0.017\pm0.004$
M$_\odot$/Mpc$^3$/yr assuming a Salpeter IMF, to which they convert
their Chabrier IMF stellar masses by adding 0.24\,dex. Removing this
correction results in the value quoted. } a total mass growth of the
red sequence of $\dot{\rho}_{A \rightarrow Q} = 0.0098$
M$_\odot$/Mpc$^3$/yr. If we assume that all galaxies that enter the
red sequence subsequently remain on the red sequence, then our results
show that $>22^{+2}_{-6}$\% and likely as much as $38^{+4}_{-11}$\% of
the growth of the red sequence at $z<1$ takes place from galaxies which
have passed through the strong post-starburst phase. If all our PSB galaxies
with \logm$>9.75$ enter the red sequence, they account for $\sim80$\%
of the growth of the red sequence. We note that these numbers are
significant, and perhaps even surprisingly high. We will return to
this point in Section \ref{sec:disc}.

The quoted errors on our mass flux include errors on the individual
galaxy stellar masses, calculated from the probability distribution
function resulting from the Bayesian SED fit, and the Poisson errors
on the QSR and SSR weights. As described in
\citet{2007A&A...474..443P}, systematic errors caused by the methods
used to estimate the masses and selection of samples lead to a typical
scatter in stellar mass densities of 0.1-0.2\,dex. If we discard our
statistical errors, and assume instead a fixed 0.15\,dex error on the
stellar mass density of the post-starburst galaxies, we find
$\dot{\rho}_{A \rightarrow Q, PSB} = 0.0038^{+0.001}_{-0.002}$
M$_\odot$/Mpc$^3$/yr. However, our comparison with the results of
\citet{2007A&A...476..137A} should be more robust due to the use of
the same stellar masses, similar survey selection criteria and survey
area.



\section{Comparison at low redshift}

\begin{figure*}
\includegraphics[scale=0.4]{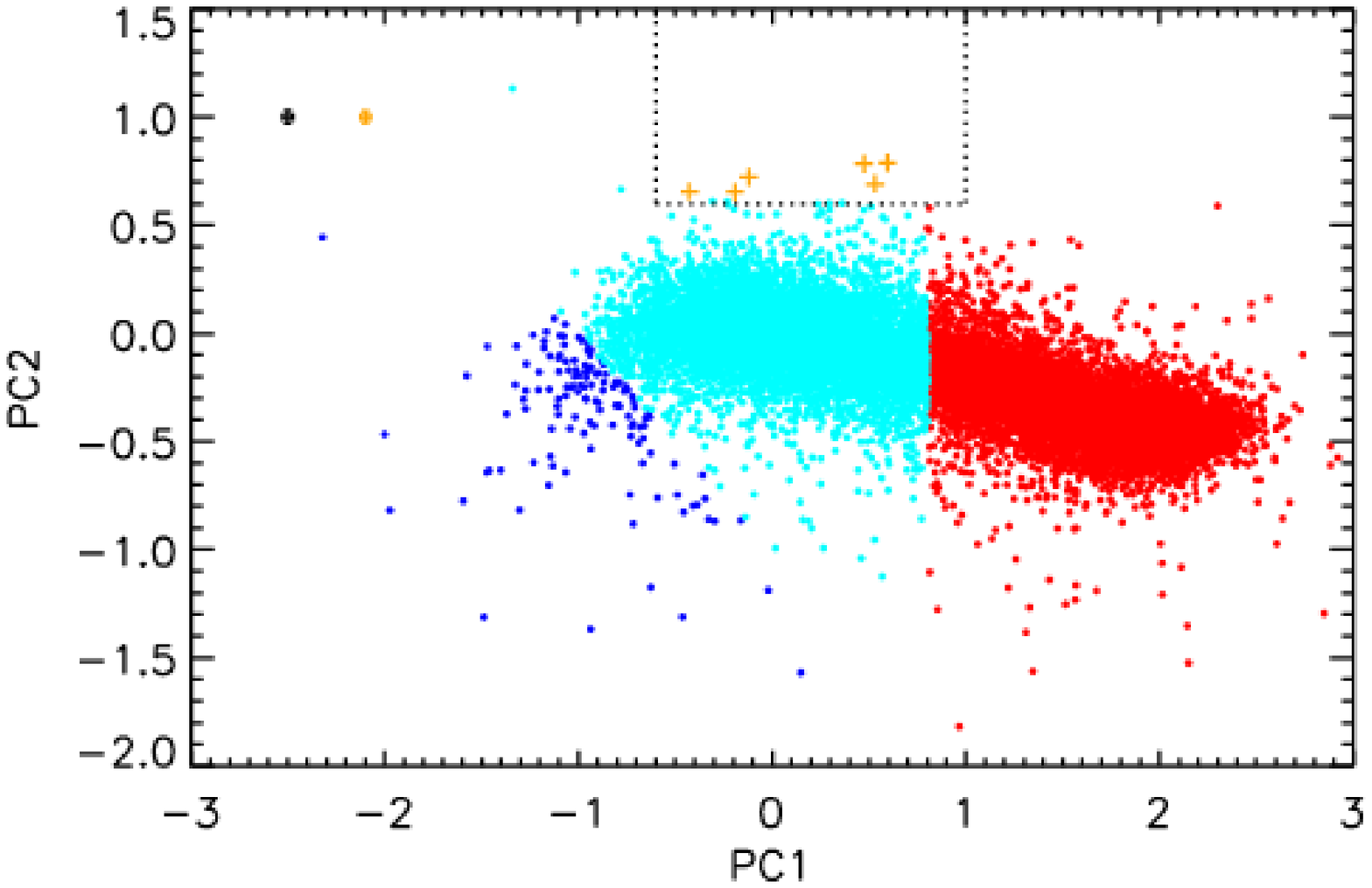}
\includegraphics[scale=0.4]{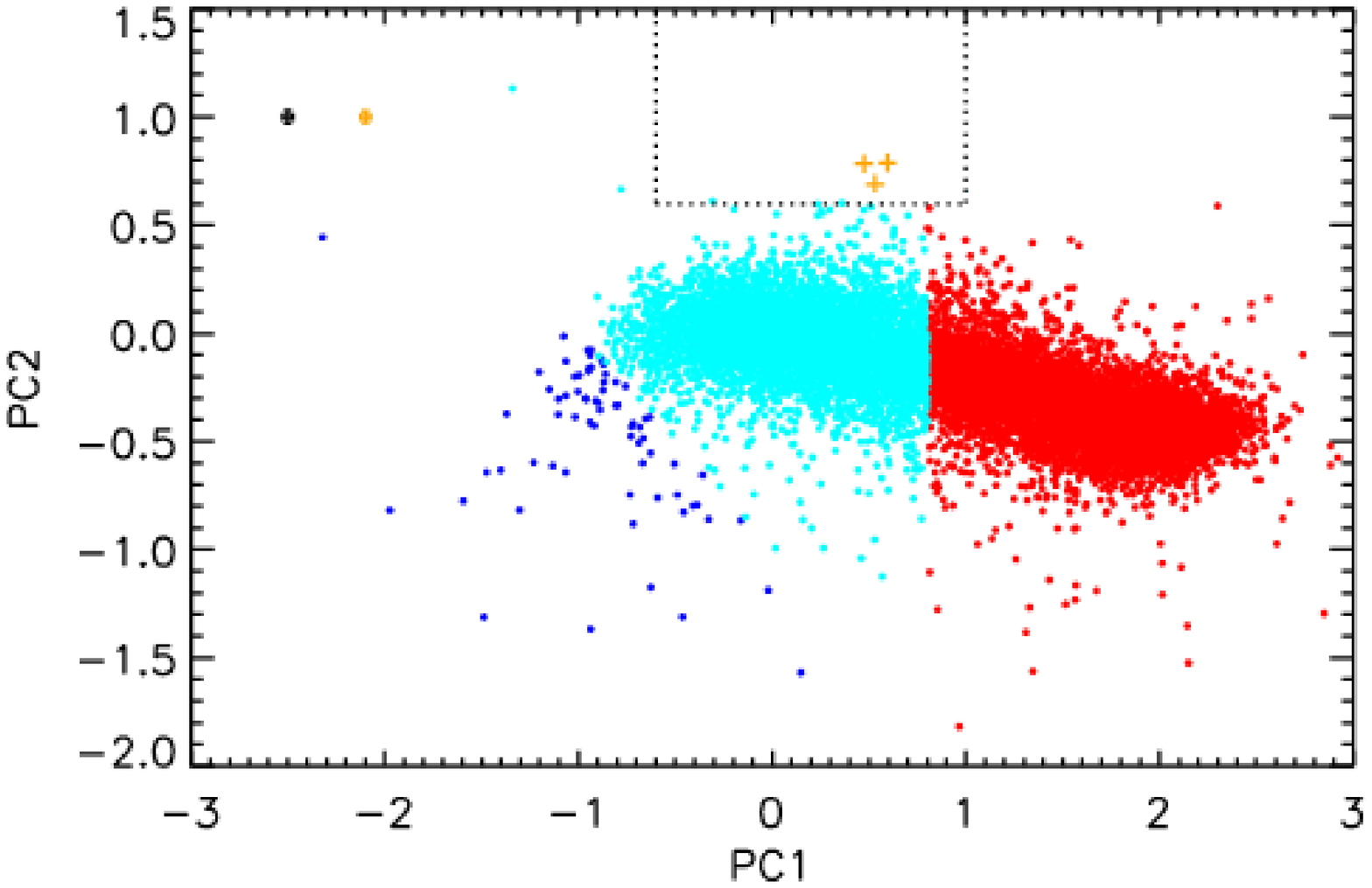}
\caption{The first two principal components (PCs) for all galaxies in
  our SDSS sample (left) and for those galaxies above our mass
  completeness limit, \logm$>9.75$ (right). For analysis
  purposes, the samples have been split into quiescent (red),
  starforming (cyan), star-bursting (blue) and post-starburst (orange)
  classes. These classes are defined in the same way as for the VVDS
  sample, except for the positioning of the partition between
  quiescent and active galaxies (see Fig. \ref{fig:pca}). The median
  errors of the whole sample (black) and the post-starburst galaxies
  alone (orange) are indicated in the top left. }\label{fig:pca_sdss}
\end{figure*}

To create a comparison sample at low redshift, we combine the Sloan
Digital Sky Survey (SDSS) spectroscopic data release 5
\citep[DR5,][]{2007ApJS..172..634A} with the UKIRT Infrared Deep Sky
Survey \citep[UKIDSS,][]{2007MNRAS.379.1599L} large area survey (LAS),
data release 3 (Warren et al. in prep.). The SDSS DR5 galaxy catalogue
\citep{2002AJ....124.1810S} covers 5740 sq. degrees and contains more
than 670,000 spectra of galaxies with a median redshift of $z\sim0.1$
and $r_{\rm AB}<17.77$. The accompanying SDSS photometric survey
provides optical $ugriz$ photometry for each object. UKIDSS is an
ongoing survey using the UKIRT Wide Field Camera
\citep[WFCAM,][]{2007A&A...467..777C} to obtain YJHK near-infrared
photometry covering 4000 sq. degrees in the same region of the sky as
the SDSS survey. The photometric system is described in
\citep{2006MNRAS.367..454H}, the pipeline processing and science
archive are described in \citep{2008MNRAS.384..637H}. The LAS reaches
a depth in $K$ of 18.4. 

For the purposes of this paper, the SDSS provides the spectroscopic
data required to locate the post-starburst galaxies, and the
near-infrared photometry of the UKIDSS survey is combined with the
optical photometry of SDSS to obtain accurate stellar
masses. Requiring the objects to be detected in all YJHK bands gives a
combined survey area of 916 sq. degrees (S. Maddox, private
communication).

\subsection{Sample selection and incompleteness corrections}
We select SDSS galaxies to have extinction corrected $r$-band
petrosian magnitudes $r_{\rm AB}<17.7$, be primary galaxy targets
({\sc primtarget = target\_galaxy} or {\sc target\_galaxy\_big} or
{\sc target\_galaxy\_red}), primary catalogue objects ($mode=1$) and
spectroscopically classified as a galaxy ({\sc specclass=2}). They are
matched to UKIDSS galaxies by identifying the closest object within
0.5\,arcsec. We remove a small fraction of galaxies (3\%) with bad
spectra by imposing a per-pixel-SNR limit of 5, and account for this
loss in our weighting scheme. Our final sample contains 14822 galaxies
with $0.05<z<0.1$.

The redshift range is selected with two things in mind. Firstly, we
require low enough redshift that our sample is complete down to
similar masses as for the VVDS survey and secondly, we require high
enough redshift to minimise the aperture bias effect caused by the 3''
SDSS fibres. This latter property of the SDSS survey means that the
spectra only probe the central regions of nearby, or massive,
galaxies, with a strong redshift dependence. For our comparative
census of post-starburst galaxies this is problematic as there is
currently no known way to robustly detect post-starburst features from
optical photometry alone. Thus, it is not possible to know for sure
whether the proceeding starburst was nuclear or global, the former
being unlikely to be connected to the build up of the red sequence. At
$z=0.05$ the fibre corresponds to a physical diameter of about 3\,kpc
($h=0.7$), which is small compared to the size of a galaxy, but is
considerably larger than the extent of a nuclear starburst
\citep{2004AJ....127..105B}. A remaining limitation caused by the
aperture bias of SDSS, is that post-starburst populations may fail to
be identified due to population gradients within the
galaxies. Assuming disks are more likely to undergo starbursts than
bulges, this would cause us to underestimate the number of PSB
galaxies.

Volume corrections and stellar masses of the galaxies are measured in
the same way as for the VVDS sample (Section \ref{sec:volume}), using
the combined UKIDSS and SDSS photometry. One final piece of
information required is the target sampling rate of SDSS.  Although
the SDSS aims for 100\% coverage in the spectroscopic catalogue, one
of a pair of neighbouring galaxies is occasionally not targeted, due
to the problem of placing fibres close together on a plate. Similarly,
densely populated regions of the sky may suffer from partial coverage
due to there being insufficient fibres available to cover the
area. For the purposes of this study, we only require the probability
that a galaxy was targeted for spectroscopic follow-up, and no further
spatial information. We use the SDSS Catalogue Archive Server (CAS) to
count the number of galaxies within SDSS spectroscopic sectors that
are targeted for spectroscopic follow-up, and the number with
spectroscopic ID numbers (i.e. that were targeted). We find that the
targeting success rate is 90\%. We calculate weights for the galaxies
by combining this value with the fraction with spectra above the SNR
limit described above.

\subsection{Calculating the Principal Components}

To calculate the Principal Components for the SDSS spectra, we first
correct for Galactic extinction, convert to air wavelengths, convolve
the spectra to VVDS resolution and finally rebin onto the eigenspectra
binning. We then proceed as for the VVDS spectra (Section
\ref{sec:indices}), to obtain the PCs for the SDSS galaxies. The
distribution of the first two components is shown in Figure
\ref{fig:pca_sdss}. The left panel shows the full sample, and the
right panel only galaxies with \logm$>9.75$. Comparing with Figure
\ref{fig:pca} there are clear differences between the SDSS and VVDS
populations. Firstly the tight blue sequence in SDSS contrasts greatly
with the cloud in the VVDS sample.  In part this is due to the smaller
errors on the spectral indices, but it may also be due to a more
``quiescent'' mode of star formation in galaxies at low redshifts. The
second noticeable difference is that the entire population is older
(larger PC1). Finally, we can see that there are fewer strong
post-starburst galaxies, despite the enormous increase in overall
sample size.

\subsection{Mass density of post-starburst galaxies at $z\sim0.07$}

\begin{figure}
\includegraphics[scale=0.4]{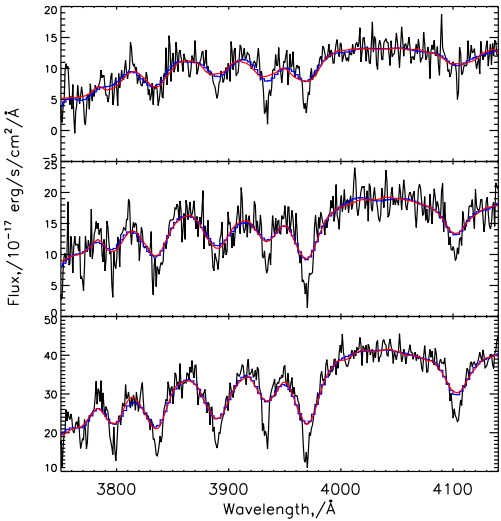}
\caption{The spectra of post-starburst galaxies culled from the SDSS
  galaxy sample. The original SDSS spectrum is shown in black, with
  the convolved spectrum in blue (histogram). The red line is the overplotted
  fit of the VVDS derived eigenspectra to the SDSS spectra. }\label{fig:spec_sdss}
\end{figure}

We identify 6 post-starburst galaxies using the same criteria as for
the VVDS sample, although from Figure \ref{fig:pca_sdss} these are
clearly only the tip of a distinct and identifiable population.  The
spectra of the three post-starburst galaxies above the mass limit of
\logm$>9.75$ are shown in Figure \ref{fig:spec_sdss}. The SDSS sample
has a high mass completeness even below this mass limit, but for the
purposes of comparison we retain the same mass limit as for the VVDS
dataset. The number density of PSBs at $z\sim0.07$ is
$5.1\times10^{-7}$, i.e. a factor of 200 lower than in the VVDS survey
at $z\sim0.7$.  For comparison, over the same redshift range the
number density of ultra-luminous infrared galaxies (ULIRGs) decreases
by a factor of $\sim$30
\citep{1998ApJS..119...41K,2004ApJ...603L..69C}.

The total completeness corrected mass of the SDSS post-starburst
galaxies with \logm$>9.75$ is $6.9^{+1.4}_{-1.3}\times10^{10}M_\odot$,
giving a mass density of $1.2\pm0.2\times 10^4M_\odot/{\rm Mpc}^3$. This is
just 0.43\% of the mass density of post-starbursts at $z\sim0.7$.

As in Section \ref{sec:massflux} an upper limit on the time the
post-starbursts are visible for is 0.6\,Gyr. Following Equation
\ref{eq:massflux} we find a mass flux of $\dot{\rho}_{A \rightarrow
Q, PSB} > 2.0\pm0.4\times10^{-5}$ M$_\odot$/Mpc$^3$/yr.  Assuming our
best estimate for the visibility of the post-starburst galaxies of
0.35\,Gyr results in  a mass flux of $\dot{\rho}_{A \rightarrow
Q, PSB} = 3.4\pm0.6\times10^{-5}$ M$_\odot$/Mpc$^3$/yr.

The SED fitting results indicate that two of the low-redshift
post-starburst galaxies have no residual star formation
(SSFR$<10^{-11}$/yr), while one has a SSFR of $10^{-10}$/yr. However,
we caution that with only the optical and NIR bands, these values are
less reliable than for the VVDS galaxies with UV-IR coverage. We have
therefore used all three galaxies to calculate the present day mass
flux through the post-starburst phase. The masses of the three SDSS
post-starburst galaxies are \logm$=$10.25, 10.15 and 10.05, similar to
the masses of the VVDS post-starburst galaxies.

We can compare our low redshift PSB mass flux to that calculated by
\citep{2007ApJS..173..342M} for galaxies in the green valley. They
assumed that green valley galaxies defined in NUV-r colours in an
SDSS/GALEX matched catalogue are all entering the red sequence
(i.e. all the mass flows from blue-to-red) to calculate a transition
mass flux of $\dot{\rho}_T = 0.033$M$_\odot$/Mpc$^3$/yr. Clearly, this
value is considerably greater (a factor of 1,000) than our
post-starburst mass flux. If the assumption that all green valley
galaxies are heading towards the red-sequence is correct, then the
post-starburst pathway to the red sequence appears truly unimportant
in the present-day Universe. Some of this discrepancy may be reduced
by appealing to the differing mass limits of the two samples. For
example, \citet{2004ApJ...602..190Q} studied a sample of
post-starburst galaxies in the SDSS, using a similar magnitude limit
to that of \citet{2007ApJS..173..342M}, which includes many lower mass
galaxies than in our sample. They measured the ``rate density'' for
post-starburst galaxies in the SDSS, i.e. the number of galaxies which
pass through the post-starburst phase, to be
$\sim4\times10^{-5}$/Mpc$^3$/Gyr. The equivalent value for our
high-mass sample is $\sim1.5\times10^{-6}$/Mpc$^3$/Gyr strongly
suggesting that the measured post-starburst mass flux would increase
were we to relax our mass limit.


\section{Discussion: are post-starburst galaxies important?}\label{sec:disc}

In this paper we have presented a sample of galaxies with strong
Balmer absorption lines and, through comparison with simulations, have
argued that these are the likely descendants of gas-rich major
mergers. The strength of their spectral features suggest that they are
post-starburst galaxies with burst mass fractions of at least 5-10\%.
However, irrespective of whether a starburst has occurred, it is
certain that their star-formation has been quenched quickly with a
timescale of $\la$0.1--0.2\,Gyr.  As discussed briefly in Section
\ref{sec:intro}, such fast quenching times are generally associated
with supernovae feedback after periods of rapid star formation induced
by galaxy collisions. An alternative scenario may be rapid fueling of
star formation in galaxies from filamentary cold-gas flows
\citep{2005MNRAS.363....2K,2008arXiv0803.4506O}, although direct
comparisons with the outputs of these simulations would be required to
test this. 


\subsection{QSOs and post-starburst galaxies}
One popular theory for the origin of the red sequence and the relation
between galaxy bulge mass and central supermassive black hole mass is
through the triggering of massive outflows driven by the central black
hole. Such outflows would quickly halt star formation and therefore
the theory suggests an evolutionary link between QSOs and
post-starburst galaxies. It is therefore of interest to compare the
number density of the post-starburst galaxies to that of QSOs at the
same redshift. Any meaningful comparison requires knowledge of the QSO
lifetime and the visibility time of the post-starburst galaxies. The
former has been estimated using a variety of methods to be $t_{\rm
QSO} \sim 10^7$ \citep[see e.g.][]{2008ApJ...676..816G} and we assume
$t_{\rm PSB} = 0.35$\,Gyr as previously. Setting
\begin{equation}
\frac{\phi_{\rm QSO}(M<M_X)}{t_{\rm QSO}} = \frac{\phi_{\rm PSB}}{t_{\rm PSB} } 
\end{equation}
we can estimate the magnitude limit of the QSOs, $M_X$, which leads to
a space density matching the observed space density of post-starburst
galaxies of $1\times 10^{-4}$/Mpc$^3$. Using the parameterised QSO
luminosity function of \citet{2004MNRAS.349.1397C} at $z=0.7$, we find
$M_X \approx M^*(z=0.7)+1.5=-22.2$ where the magnitudes are
$b_j$--band and Vega zero-point. That the number density of strong
post-starburst galaxies coincides with that of moderately powerful
QSOs, 1.5 magnitudes fainter than $M^*$, does not of course
necessarily imply an evolutionary link. The AGN luminosity function
continues to rise to fainter magnitudes \citep{2005AJ....129.1795H}
and uncovering the triggering mechanisms of AGN as a function of
luminosity remains one of the biggest observational challenges for
extragalactic astronomy.

The number density of the post-starburst galaxies evolves
rapidly. Extrapolating the evolution of the QSO luminosity function of
\citet{2004MNRAS.349.1397C} to $z=0.07$, and taking $t_{\rm QSO} \sim
10^7$ as before, we find a QSO number density of $3.1\times
10^{-5}$Mpc$^{-3}$ above the same magnitude limit of $M_X=-22.2$, a
factor of a few below the $z=0.7$ number density. So current results
suggest that the evolution in number density of QSOs is not as strong
as the evolution in the PSB number density. To our knowledge, an
observed QSO luminosity function at $z\sim0$ is not currently
available, and would be required for a more detailed comparison.

\subsection{Major mergers and post-starburst galaxies}
If, as we have suggested, the strong post-starburst galaxies selected
in this paper are the result of major mergers, it is instructive to
compare the number of post-starburst galaxies to the number of close
galaxy pairs in the same survey volume. \citet{2008arXiv0807.2578D}
measure the fraction of close spectroscopic pairs to derive the merger
rate using the same VVDS galaxy sample as in this paper. For galaxies
with \logm$>10$ they derive a merger rate of
$2\times10^{-4}$Mpc$^{-3}$Gyr$^{-1}$, with a magnitude difference
criteria such that they are sensitive to mergers with mass ratios
$\ga$4:1. This can be compared with the number of post-starburst
galaxies, divided by the time during which they are detectable
i.e. $2.9\times10^{-4}$Mpc$^{-3}$Gyr$^{-1}$ for $t_{\rm
PSB}=$0.35\,Gyr. Given the small number statistics of both samples,
uncertainty in both merger and PSB timescales and the problem of the
unknown magnitude difference between the PSB progenitors, it is
encouraging that these numbers are so close. Of course, larger samples
and further detailed observations of the post-starburst galaxies will
greatly aid our understanding of their origin and possible link to
gas-rich major mergers. As with the QSO number density, the current
estimates of the evolution in merger rate with redshift show less
evolution than observed in the PSB number density. Although
\citet{2008arXiv0807.2578D} find that the evolution with redshift is
stronger for lower luminosity and lower mass galaxies, with the merger
rate decreasing by a factor of two between a redshift of 1 and 0.5,
this is still not as strong as the evolution in PSB number
density. Finally, we note that out of the 36 close galaxy pairs
identified by \citet{2008arXiv0807.2578D} one galaxy is part of the
post-starburst sample of this paper. Given the small probability of a
chance coincidence, this suggests that in at least one case the
post-starburst stellar population is linked with tidal disruption
caused by an ongoing major merger.

The strong evolution in PSB number (and mass) density compared to the
evolution of both QSO number density and galaxy major merger rate,
leads us to the question of what may cause such a sharp decline in the
number of post-starburst galaxies. Through our comparison with merger simulations,
we have identified two important factors in creating a post-starburst
galaxy: gas mass fraction and timescale of the starburst. Larger gas
mass fractions provide more fuel for the starburst, leading to
stronger and more prolonged post-starburst signatures. Because gas is
used-up in the formation of stars, it is reasonable to assume that gas
mass fraction decreases with decreasing redshift, a fact that may play
a leading role in the global decrease of star formation rate density
since $z\sim1$. A second effect is the starburst timescale, which is
generally assumed to be linked to the disk dynamical timescale, which
in turn is linked to the dynamical time of the halo. This latter value
is known to increase with redshift, by a factor of 2 between a
redshift of 1 and 0 \citep[E. Neistein priv.
comm.,][]{2001PhR...349..125B}. A lengthening of the duration
of the starbursts, together with a decreasing amount of fuel
available, may be the dominant mechanism responsible for the decrease
in number density of post-starburst galaxies. Clearly further
simulations will be required to test these ideas.

\subsection{The environments of PSB galaxies}
Given the possible link of PSB galaxies to gas-rich major mergers, it
is interesting to investigate the local environments of the
galaxies. \citet{2006A&A...458...39C} measured the density on 5 and
8$h^{-1}$Mpc scales around galaxies in the VVDS survey. Of the 16 PSB
galaxies with \logm$>9.75$, 15 have available density measures. A
control sample of 2122 VVDS galaxies with good quality redshift flags
was selected for comparison. We do not detect any significant
difference in the mean or median densities around the PSB galaxies, in
agreement with the results of \citet{2008arXiv0805.0004Y}. It is,
however, interesting to see that the spread in local density values
for the PSB sample is very large. In detail, 3 lie in underdense
environments ($\delta_5<0$), 9 in normally overdense environments
($0<\delta_5<1$, while the median value for the control sample is
$\delta_5 = 0.44 \pm 0.02$) and 3 in strongly overdense environments
($\delta_5>1$).  Furthermore, there is no clear trend between
star-forming vs. non-star-forming PSB galaxies. The presence of PSB
galaxies in all types of environment will be an important constraint
for understanding their origin and evolution.

\subsection{Building the red sequence through post-starburst galaxies}
In Section \ref{sec:massflux} we compared the mass flux through the
post-starburst phase to the mass build--up of the red sequence as
measured by \citet{2007A&A...476..137A}. We found a value of
$\sim$40\% for PSBs with no residual star formation, or 80\% for all
PSBs with \logm$>9.75$. These numbers, although consistent (i.e. not
greater than 100\%), perhaps still appear surprisingly high if we
expect other quenching mechanisms with slower timescales to also play
a role in moving mass onto the red sequence
\citep{2008MNRAS.387...79V}. The key question is whether there is a
one--way flow of galaxies from the blue to the red sequence?  What
prevents a galaxy from re-starting star formation due to subsequent
inflow of gas after experiencing a major merger? SPH simulations of
galaxy mergers invoke strong mechanical AGN feedback to prevent
subsequent star formation and allow the galaxy to remain on the
red-sequence. The large amount of mass flowing to the red sequence
through the post-starburst phase at high redshift, may in fact cause
us to question the fact that strong mechanical feedback is effective
in the long-term for the majority of galaxies. Semi-analytic
cosmological models provide one method to test galaxy formation
scenarios in a full cosmological context and could help to understand
the directions of mass fluxes. A full comparison of the recent star
formation histories of the VVDS and SDSS galaxies presented in this
paper with semi-analytic models is underway. A full census of
different types of transition galaxies as a function of redshift, when
combined with accurate measurements of the mass densities on the blue
and red sequence, will help to reveal the true importance of
mechanical gas expulsion mechanisms (feedback) on the global evolution
of the galaxy population.

We have found that strong post-starburst galaxies have stellar masses
  similar to those of the least massive galaxies on the red
  sequence. This has important implications for our understanding of
  the physical processes involved in the build-up of the red sequence,
  and in particular the subsequent role of dry-mergers in forming the
  shape of the red sequence mass function observed today
  \citep{2006ApJ...640..241B,2006ApJ...636L..81N}.


\section{Summary}

Using a PCA analysis of the spectra of the VVDS deep spectroscopic
galaxy survey, we have selected a sample of 16 galaxies with strong
Balmer absorption lines at $0.5<z<1.0$ and above a mass completeness
limit of \logm$>9.75$. Through comparison with a suite of SPH merger
simulations and toy starburst models we have shown that these galaxies
are likely to have undergone a strong starburst within the last few
tenths of a Gyr, with burst mass fractions of order 10\%, similar to
those derived for LIRGS at these redshifts
\citep{2006A&A...458..369M}.  The key requirement for the observation
of the post-starburst galaxies, is a fast quenching timescale of
$\la0.1-0.2$\,Gyr. We show that there is a maximum visible lifetime
for the post-starburst galaxies of 0.6\,Gyr, but for lower gas mass
fractions of $\sim$20-40\% a lifetime of 0.35\,Gyr is more likely.

The key results of this paper are:
\begin{itemize}
\item {\bf Number density:} PSB galaxies with \logm$>9.75$ have a
  number density of $10^{-4}$ per Mpc$^3$. Assuming they are visible
  for an average of 0.35\,Gyr, and that a QSO shines for $10^7$ years,
  this is equal to the number density of QSOs brighter than
  $M^*(z=0.7)+1.5$ \citep{2004MNRAS.349.1397C}.
\item{\bf Masses and mass density:} Summing all the mass in the
  post-starburst population gives a mass density of $\log(\rho^*/{\rm
  M_\odot/Mpc}^3) = 6.43_{-0.04}^{+0.06}$. The mass distribution of
  the PSB galaxies rises to the mass completeness limit of
  \logm$>9.75$, therefore measuring the complete mass distribution for
  post-starburst galaxies at $z\sim0.7$ will require a deeper
  survey. The true mass distribution is crucial for understanding the
  relative importance of different mechanisms for causing
  post-starburst galaxies \citep{2007MNRAS.382..960K}.
\item {\bf Mass flux:} We select the 5 PSB galaxies with no residual
  star formation according to multiwavelength SED fitting. Taking
  0.6\,Gyr as an upper limit on the visibility time of the PSB
  galaxies gives a lower limit on the mass flux of $\dot{\rho}_{A
  \rightarrow Q, PSB} > 0.0022^{+0.0002}_{-0.006}$
  M$_\odot$/Mpc$^3$/yr. Taking our best estimate for $t_{\rm PSB}$ of
  0.35\,Gyr gives $\dot{\rho}_{A \rightarrow Q, PSB} =
  0.0038^{+0.0004}_{-0.001}$ M$_\odot$/Mpc$^3$/yr. Comparing this to
  the rate of build up of the red sequence
  \citep{2007A&A...476..137A}, we find $38^{+4}_{-11}$\% of the growth
  of the red sequence at $z<1$ takes place from galaxies which have
  passed through the strong post-starburst phase, assuming all these
  galaxies subsequently remain on the red sequence.
\item {\bf Environment:} We use the local density on $5h^{-1}$Mpc scales
  from \citet{2006A&A...458...39C} to show that PSB galaxies are found
  in all environments, from underdense to strongly overdense.
\item {\bf Low vs. High $z$:} We compare our high-redshift results to
  a sample of galaxies with $0.05<z<0.1$ selected from the SDSS and
  UKIDSS surveys. Despite the enormous increase in sample size, only 3
  post-starburst galaxies of the same strength as in the VVDS sample
  are detected for the same mass limit. We find the mass density of
  strong post-starburst galaxies decreases by a factor of 230 and
  number density by a factor of 200 from $z\sim0.7$ to $z\sim0.07$
  \citep[see also][]{1997ApJ...481...49H,2006ApJ...642...48L}.
\end{itemize}

This paper finds that post-starburst galaxies, although rare in the
local Universe, are of global importance at higher redshift. With
larger and deeper spectroscopic surveys, exciting new constraints
could be obtained on the processes which drive galaxy evolution. The
joint analysis of imaging and spectroscopy will help to break
observational degeneracies, especially when combined with
``observations'' of galaxy simulations. On the advent of an era of
large broad band photometric surveys, it is important not to forget
the wealth of additional information on galaxy evolution available
from the additional investment in spectroscopy.


\section*{Acknowledgments}

The authors would like to thank Celine Eminian and Steve Warren for
their help with using the UKIDSS survey; Henry McCracken for all his
help with the VVDS and CFHTLS surveys and the photo-zs, and Olivier
Ilbert for aiding our understanding of the photometric catalog
generation; Olga Cucciati for providing the density measurements for
the VVDS galaxies; Christy Tremonti, Crystal Martin, John Silverman
and Eyal Neistein for interesting discussions; the MAGPop Network and
especially PI Guinevere Kauffmann for providing the networking
opportunity that led to the start of this project; and the anonymous
referee for their careful reading of the manuscript and constructive
comments for improvement. VW acknowledges an Institut d'Astrophysique
EARA visitor grant and the Aspen center for Physics for providing the
quiet surroundings in which this paper was mostly written. During this
project, VW and CJW were supported by the EU MAGPop Marie Curie
network. AP was financed by the research grant of the Polish Ministry
of Science PBZ/MNiSW/07/2006/34A.

The numerical simulations were performed on the Munich Observatory
SGI-Altix 3700 Bx2 machine, which was partly funded by the Cluster of
Excellence: ``Origin and Structure of the Universe''


\end{document}